\documentclass{pasj02}
\Received{$\langle$reception date$\rangle$}
\Accepted{$\langle$acception date$\rangle$}
\Published{$\langle$publication date$\rangle$}

\usepackage[version=4]{mhchem}
\usepackage{adjustbox}
\usepackage{caption}
\usepackage{xcolor}

\begin{document}

\title{Detailed analysis of multi-line molecular distributions
in the Seyfert galaxy NGC 1068: Possible effect of
the AGN outflow to the starburst ring}
\author{Hiroma Okubo$^1$, Toshiki Saito$^2$, Shuro Takano$^3$, Nario Kuno$^{1,4}$, Akio Taniguchi$^{5, 6}$, Taku Nakajima$^7$, Nanase Harada$^{8, 9}$ and Ken Mawatari$^{10}$}%

\altaffiltext{1}{Division of Physics, Faculty of Pure and Applied Sciences, University of Tsukuba, Tsukuba, Ibaraki 305-8571, Japan}
\altaffiltext{2}{Faculty of Global Interdisciplinary Science and Innovation, Shizuoka University, 836 Ohya, Suruga-ku, Shizuoka 422-8529, Japan}
\altaffiltext{3}{Department of Physics, General Studies, College of Engineering, Nihon University, Tamura-machi, Koriyama, Fukushima 963-8642, Japan}
\altaffiltext{4}{Tomonaga Center for the History of the Universe, University of Tsukuba, 1-1-1 Tennodai, Tsukuba, Ibaraki 305-8571, Japan}
\altaffiltext{5}{Kitami Institute of Technology, 165 Koen-cho, Kitami, Hokkaido 090-8507, Japan}
\altaffiltext{6}{Department of Physics, Graduate School of Science, Nagoya University, Furo-cho, Chikusa-ku, Nagoya, Aichi 464-8602, Japan}
\altaffiltext{7}{Institute for Space-Earth Environmental Research, Nagoya University, Furo-cho, Chikusa-ku, Nagoya, Aichi 464-8601, Japan}
\altaffiltext{8}{National Astronomical Observatory of Japan, 2-21-1 Osawa, Mitaka, Tokyo 181-8588, Japan}
\altaffiltext{9}{Astronomical Science Program, Graduate Institute for Advanced Studies, SOKENDAI, 2-21-1 Osawa, Mitaka, Tokyo 181-1855, Japan}
\altaffiltext{10}{Waseda Research Institute for Science and Engineering, Faculty of Science and Engineering, Waseda University, 3-4-1, Okubo, Shinjuku-ku, Tokyo 169-8555, Japan}

\email{s2430049@u.tsukuba.ac.jp}

\KeyWords{galaxies: individual (NGC 1068) --- methods: data analysis --- methods: statistical --- galaxies: ISM --- radio lines: galaxies}

\maketitle

\begin{abstract}
We apply principal component analysis (PCA) to the integrated intensity maps of 13 molecular lines of the nearby type-2 Seyfert galaxy NGC 1068 obtained by Atacama Large Millimeter/sub-millimeter Array (ALMA) to objectively visualize the features of its center, (1) within a radius of about 2 kpc ($\sim$ 27".5; hereafter the "overall region") and (2) the ring shaped starburst region between 750 pc ($\sim$ 10") and 2 kpc ($\sim$ 27".5) of the galaxy (hereafter the "SB ring region").
PCA is a powerful unsupervised machine learning technique that extracts key information through dimensionality reduction. 
The PCA results for the overall region have a possibility to reconstruct a map representing the approximate H$_2$ column density and difference of volume density and/or chemical composition between the circumnuclear disk (CND) and the starburst ring (SB ring).
Additionally, the PCA results for the SB ring region have a possibility to reconstruct a map representing the approximate H$_2$ column density and distinction between starburst dominated region and shock dominated region.
Furthermore, the PCA results for the SB ring region indicate a possible interaction between the Active Galactic Nucleus (AGN) outflow and gas in the SB ring. Although further investigation is required, we suggest that the AGN outflow interacts with gas in the SB ring, as this feature is consistent with the direction of the AGN outflow and is contributed by CN, C$_2$H and HCN, which are known to be enhanced by the AGN outflow.
These results demonstrate that PCA can effectively extract features even for galaxies with complex structures, such as AGN + SB ring. This study also implies that PCA has the potential to uncover previously unrecognized phenomena by visualizing latent structures in multi-line data.

\end{abstract}

\section{Introduction}

\begin{figure*}[htbp]
\hspace{0.5cm} 
\includegraphics[width=16cm]{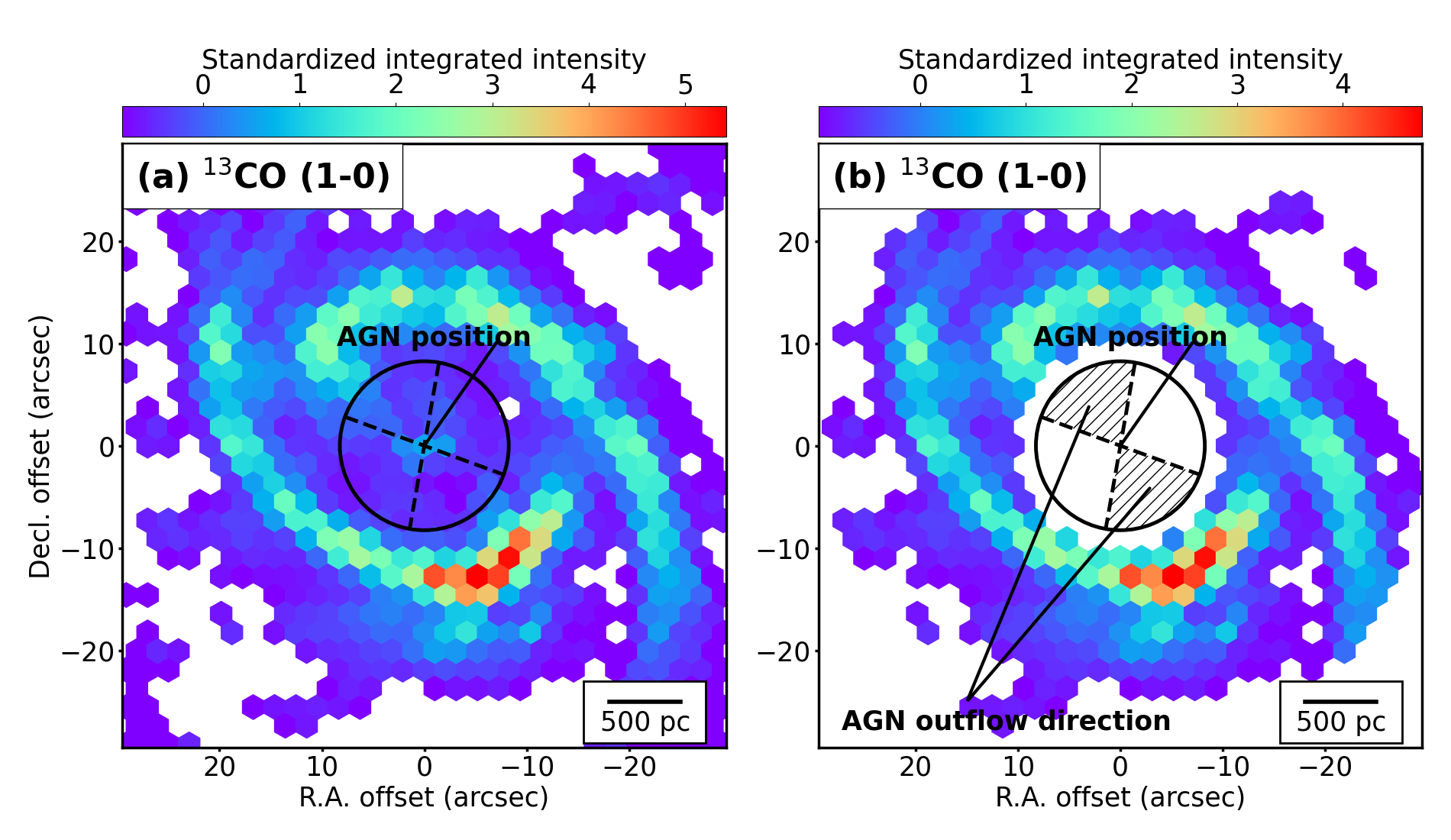}
\caption{
The $^{13}$CO(1–0) standardized integrated intensity maps of (a) overall region and (b) SB ring region of NGC 1068 used for PCA. The two dashed lines intersecting the AGN position (\cite{Roy et al. 1998}) indicate the approximate outer boundaries of the ionized gas cones (\cite{Mingozzi et al. 2019}). The central coordinates of this image are ($\alpha$, $\delta$)$_{J2000}$ = (2$^h$42$^m$40$^s$.7132, -0$^{\circ}$0$^{\prime}$47$^{\prime\prime}$.655). The black circle marks the field of view (FoV) of the [CI] data (\cite{Saito et al. 2022a}), which covers nearly the central 1 kpc. The diagonal lines enclosed by the black circle and dashed lines represent the AGN outflow direction (e.g., \cite{Das et al. 2006}; \cite{Saito et al. 2022}).\\
{Alt text: Example of standardized integrated intensity maps labeled as (a) and (b). Both the x axis and y axis are in units of "arcsec". Map (a) does not mask the central region ($\sim$ 10"), while map (b) masks the central region ($\sim$ 10").}
}
\label{13co_maps}
\end{figure*}

\begin{figure*}[htbp]
\centering
\includegraphics[width=13cm]{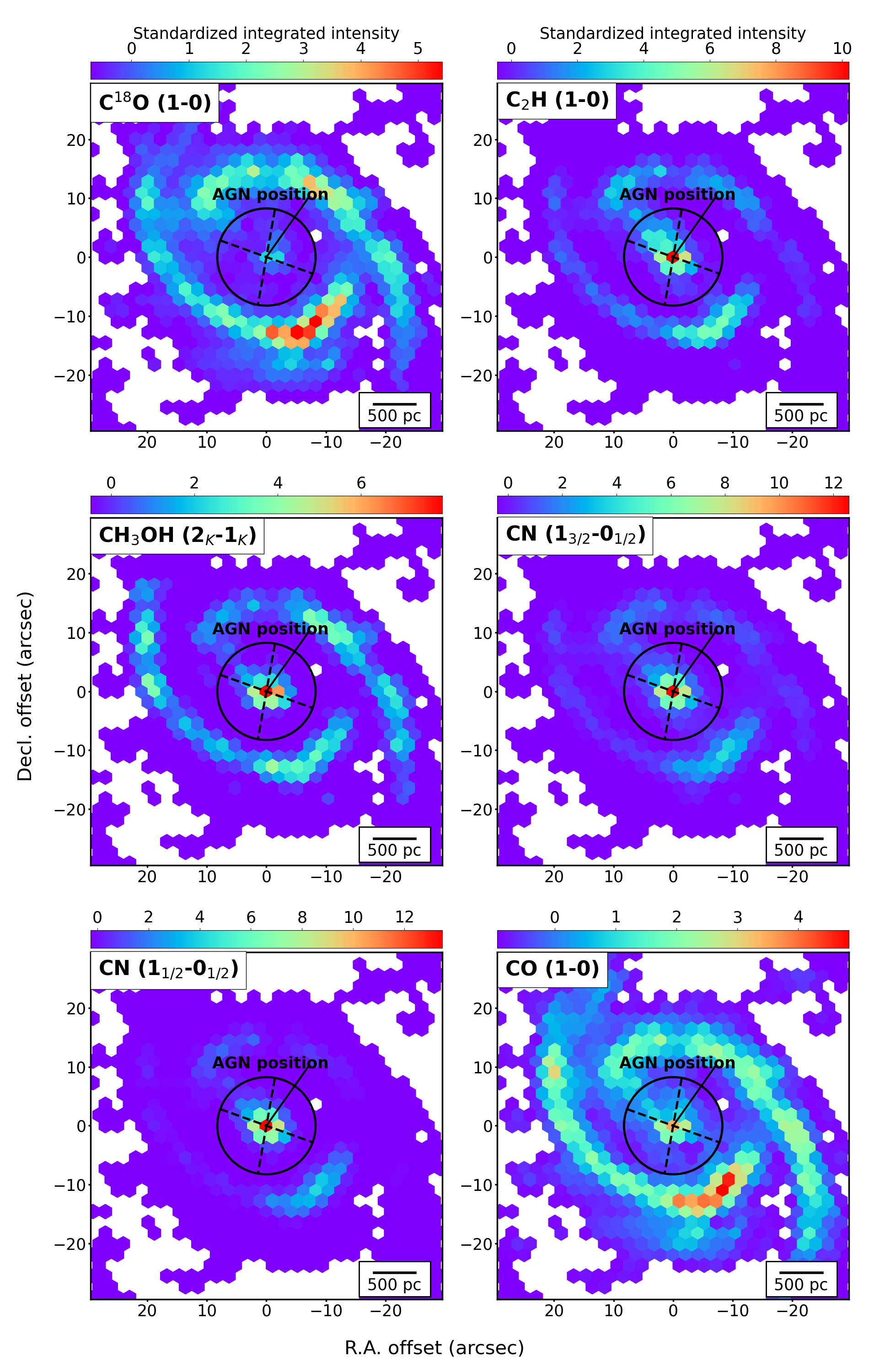}
\caption{
The standardized integrated intensity maps usd for the PCA$_{\rm OA}$. The molecular line name is shown in the top left corner of each panel. The AGN position, the two dashed lines and the black circle are same as Figure \ref{13co_maps}.\\
{Alt text: Standardized integrated intensity maps of six molecular lines. The central region is not masked, similar to Figure \ref{13co_maps} (a).}
}
\label{mom0_whole_1}
\end{figure*}

\begin{figure*}[htbp]
\centering
\includegraphics[width=13cm]{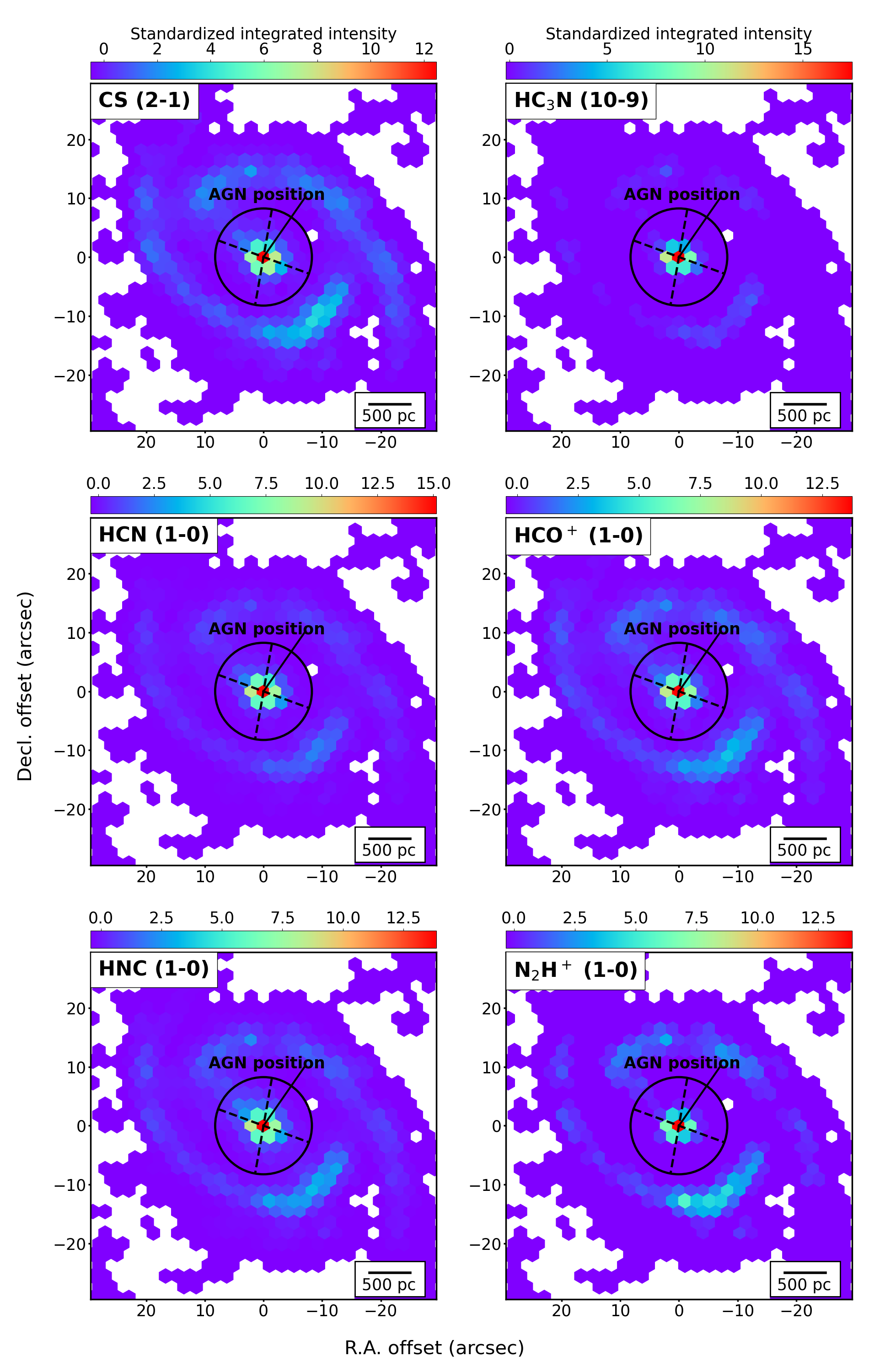}
\caption{As same as Figure \ref{mom0_whole_1}.\\
{Alt text: Standardized integrated intensity maps of another six molecular lines. The central region is not masked, similar to Figure \ref{13co_maps} (a).}
}
\label{mom0_whole_2}
\end{figure*}

\begin{figure*}[htbp]
\centering
\includegraphics[width=13cm]{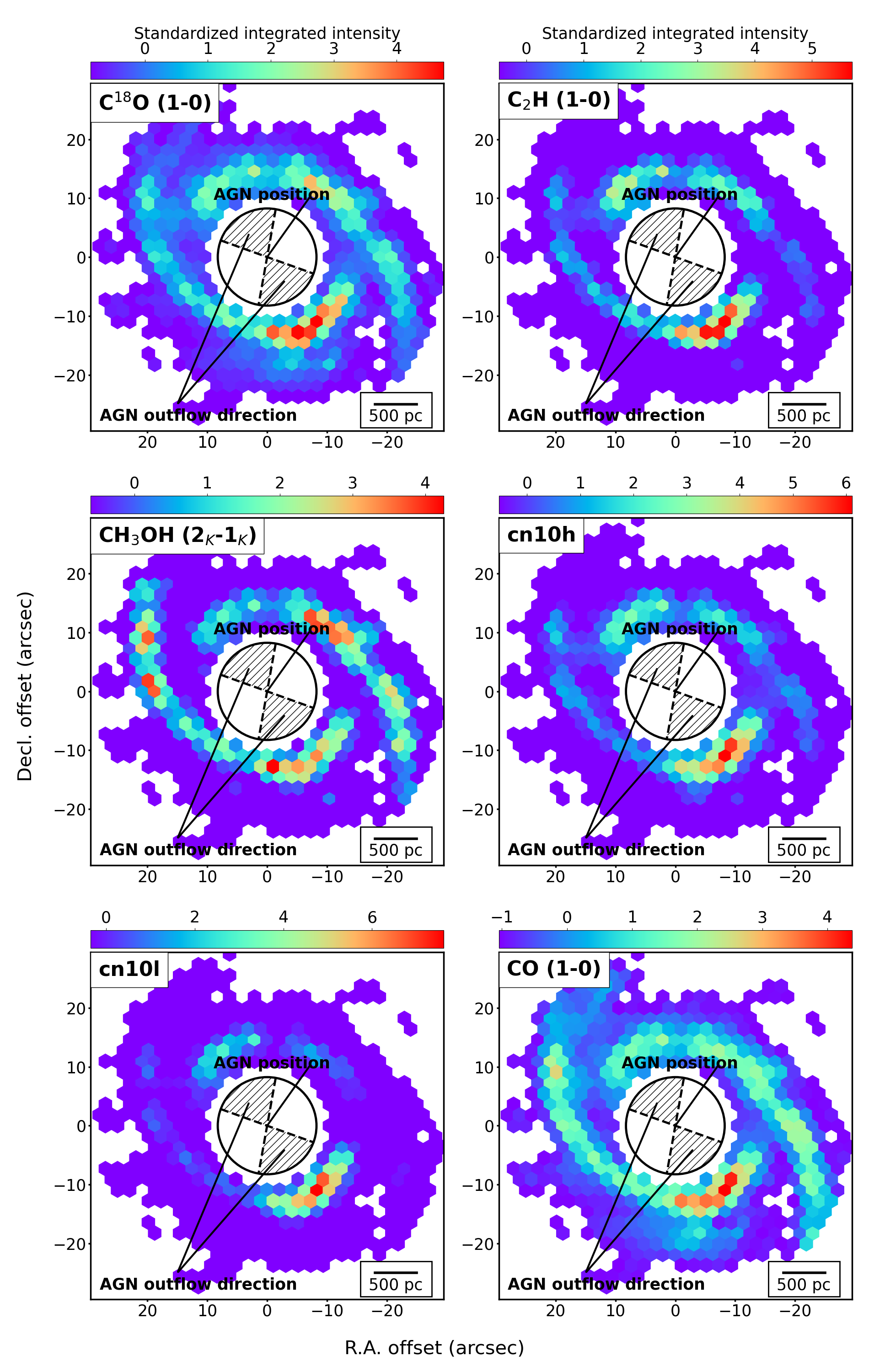}
\caption{
The standardized integrated intensity maps usd for the PCA$_{\rm SB}$. The molecular line name is shown in the top left corner of each panel. The AGN position, the AGN outflow direction, the two dashed lines and the black circle are same as Figure \ref{13co_maps}.\\
{Alt text: Standardized integrated intensity maps of six molecular lines. The central region is masked, similar to Figure \ref{13co_maps} (b).}
}
\label{mom0_sbr_1}
\end{figure*}

\begin{figure*}[htbp]
\centering
\includegraphics[width=13cm]{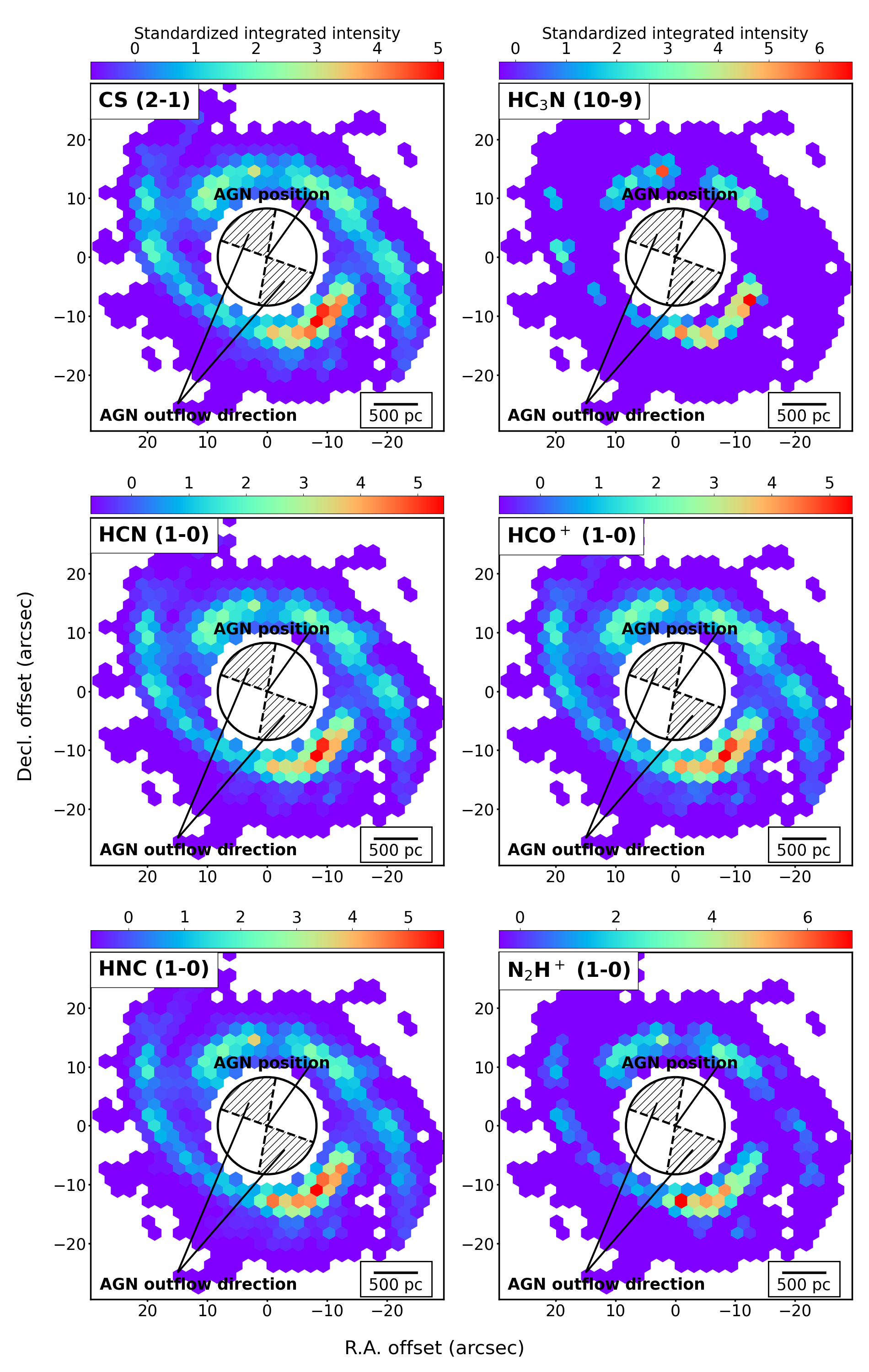}
\caption{As same as Figure \ref{mom0_sbr_1}.\\
{Alt text: Standardized integrated intensity maps of another six molecular lines. The central region is masked, similar to Figure \ref{13co_maps} (b).}
}
\label{mom0_sbr_2}
\end{figure*}

\begin{figure}[htbp]
\centering
\includegraphics[width=8cm]{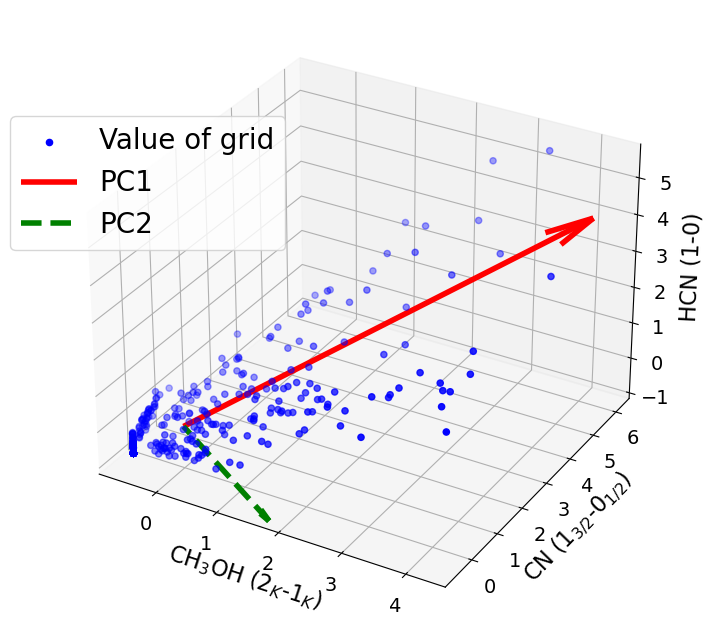}
\caption{
An image showing an example of PCA in three dimensions.\\
{Alt text: Illustrative diagram of PCA applied to the standardized integrated intensity maps.}
}
\label{PCA_outline map}
\end{figure}

Star formation is a key driver of galaxy evolution, and molecular clouds are the birthplaces of stars. 
The chemical composition within molecular clouds is known to vary significantly depending on the physical processes occurring there. For instance, HC$_3$N and N$_2$H$^+$ intensities increase in high-density regions because their critical densities ($n_\mathrm{crit}$) are about 10$^5$ $\mathrm{cm}$$^{-3}$ (e.g., \cite{Kauffmann et al. 2017}; \cite{Tanaka et al. 2018}).
Additionally, it is considered that CN and C$_2$H are produced by the dissociation of CO and HCN in UV environments (e.g., \cite{Fuente et al. 1993}; \cite{Boger and Sternberg 2005}; \cite{GarcíaBurillo et al. 2017}; \cite{Saito et al. 2022}). 
Furthermore, molecules such as CH$_3$OH and HNCO are thought to sublimate from dust surfaces, and are then detected in their gas phase in warm regions (e.g., \cite{Garay and Lizano. 1999}; \cite{Watanabe et al. 2003}; \cite{Rodr\'{\i}guez-Fern\'{a}ndez et al. 2010}; \cite{Xie et al. 2023}). 
Thus, investigating the chemical composition provides critical insights into the physical and chemical processes occurring in galaxies.
This approach, known as chemical diagnostics, is essential for studying galaxy evolution (e.g., \cite{Harada et al. 2019}).
Thanks to advances in single-dish radio telescopes and the Atacama Large Millimeter/submillimeter Array (ALMA), chemical diagnostics has become increasingly important. However, the interpretation of these data has also become more complex due to the growing volume of information.

With the increasing volume of data, unsupervised machine learning techniques such as Principal Component Analysis (PCA), which excel in dimensionality reduction and feature extraction, have been widely adopted in astronomical studies.
For example, \citet{Donnan et al. 2024} applied PCA to the integral field unit (IFU) data of polycyclic aromatic hydrocarbons (PAH) emission with the James Webb Space Telescope (JWST).
As a result, they successfully extracted information related to galaxy rotation.
There are also studies that have applied PCA to integrated intensity maps (e.g., \cite{Meier and Turner 2005}; \cite{Okoda et al. 2021}; \cite{Saito et al. 2022}; \cite{Harada et al. 2024}).
\citet{Harada et al. 2024} applied PCA to integrated intensity maps of NGC 253 obtained with ALCHEMI (ALMA Comprehensive High-resolution Extragalactic Molecular Inventory), an ALMA large program that aims to produce the most comprehensive catalog of extragalactic molecules in the central molecular zone of NGC 253 (e.g., \cite{Martin et al. 2021}; \cite{Huang et al. 2023}; \cite{Tanaka et al. 2024}), and successfully extracted valuable information from 150 integrated intensity maps.
These studies suggest that PCA can reconstruct data in a manner that reflects distinct physical phenomena.

\begin{table*}[h]
\centering
\caption{Line properties and imaging properties}
\begin{tabular}{cc|ccc|ccc}
   \hline
     &  & & Overall region  & & &  SB ring region\\ \hline
    Line & $\nu_{\mathrm{rest}}$ & Med $\sigma_{\mathrm{hex}}$ & S/N$_{\mathrm{hex}}$ & $N_{\mathrm{hex}}$ & Med $\sigma_{\mathrm{hex}}$ & S/N$_{\mathrm{hex}}$ & $N_{\mathrm{hex}}$ \\ \hline
   -- & (GHz) & (K km/s)  & (16th-50th-84th-100th) & -- & (K km/s)  & (16th-50th-84th-100th) & -- \\
   \hline\hline
   C$_2$H ($N$=1-0) & 87.316898 & 0.49 & 4.8-7.6-12.5-22.9 & 160 & 0.45 & 4.9-7.6-12.4-22.9 & 127 \\
   HCN ($J$=1-0) & 88.631602 & 0.80 & 5.8-10.6-23.1-124.2 & 171 & 0.89 & 5.7-10.4-21.8-57.1 & 166 \\
   HCO$^+$($J$=1-0) & 89.188525 & 0.97 & 5.1-8.8-17.7-57.1 & 194 & 1.08 & 5.1-8.6-17.2-42.1 & 165 \\
   HNC ($J$=1-0) & 90.663568 & 0.25 & 6.5-13.0-32.3-146.5 & 176 & 0.26 & 6.4-12.5-30.7-93.1 & 147 \\
   HC$_3$N ($J$=10-9) & 90.979023 & 0.12 & 4.7-6.6-12.8-51.4 & 61 & 0.11 & 4.6-6.5-9.3-17.1 & 47 \\
   N$_2$H$^+$ ($J$=1-0) & 93.173977 & 0.20 & 4.8-7.9-15.3-37.3 & 124 & 0.19 & 4.8-7.9-13.6-33.0 & 110 \\
   CH$_3$OH ($J$=2$_{\it K}$-1$_{\it K}$) & 96.744550 & 0.17 & 5.3-9.1-15.2-28.4 & 179 & 0.15 & 5.3-9.0-14.9-28.4 & 151 \\
   CS ($J$=2-1) & 97.980953 & 0.33 & 5.5-10.1-20.5-74.2 & 202 & 0.33 & 5.5-10.1-19.8-49.1 & 167 \\
   C$^{18}$O ($J$=1-0) & 109.782173 & 0.26 & 5.3-10.2-20.0-40.2 & 223 & 0.28 & 5.6-10.8-20.2-40.2 & 164 \\
   $^{13}$CO ($J$=1-0) & 110.201354 & 0.57 & 6.4-13.4-34.9-99.3 & 220 & 0.57 & 6.9-15.7-37.0-99.3 & 175 \\
   CN ($N$=1$_{1/2}$-0$_{1/2}$) & 113.191279 & 0.33 & 5.0-8.0-18.3-47.0 & 109 & 0.32 & 4.9-7.6-16.1-37.2 & 98 \\
   CN ($N$=1$_{3/2}$-0$_{1/2}$),  & 113.490970 & 0.98 & 5.0-7.9-15.7-44.6 & 171 & 0.95 & 5.0-7.7-13.5-44.6 & 142 \\
   CO ($J$=1-0) & 115.271202 & 3.41 & 10.1-28.9-69.6-196.1 & 240 & 3.40 & 11.5-33.6-78.2-196.1 & 183 \\
   \hline
\end{tabular}
\caption*{\textbf{Notes.} Line: line name. $\nu_{\mathrm{rest}}$: rest frequency of line (\cite{Endres et al. 2016}). Med $\sigma_{\mathrm{hex}}$: median noise rms of the detected hexagons. S/N$_{\mathrm{hex}}$: S/N distribution of the detected hexagons. $N_{\mathrm{hex}}$: number of detected hexagons.}
\label{line table}
\end{table*}

In this study, we apply PCA to several integrated intensity maps of NGC 1068. 
NGC 1068 is a nearby type-2 Seyfert galaxy with both an Active Galactic Nucleus (AGN) and a starburst (SB) ring. It exhibits rich molecular lines and chemical variations (e.g., \cite{Wang et al. 2014}; \cite{Takano et al. 2014}; \cite{Aladro et al. 2015}). A near-infrared bar has also been observed and substructures have been identified in molecular lines and continuum (e.g., \cite{Scoville et al. 1988}; \cite{Scourfield et al. 2020}; \cite{Sanches Garcia et al. 2022}).
We have two objectives for applying PCA to NGC 1068.
The first objective is to assess the validity of PCA for feature extraction in galaxies that host both the AGN and the SB ring. While PCA has been applied to specific systems such as starburst galaxies like NGC 253 and the AGN-dominated central region of NGC 1068, its effectiveness for more complex galaxies containing both the AGN and the SB ring has not been thoroughly examined. Therefore, we apply PCA to molecular lines of the central $\sim$ 2 kpc ($\sim$ 27".5) of NGC 1068 (hereafter overall region), which includes both the AGN and the SB ring, to evaluate its capability in extracting meaningful features from such complex systems.
The second objective is to investigate the potential influence of the AGN outflow on the SB ring. Previous studies, such as \citet{Saito et al. 2022}, have confirmed interactions between the AGN outflow and gas within the central $\sim$ 1 kpc of NGC 1068. Additionally, \citet{Nakajima et al. 2023} suggested that the AGN outflow might extend beyond the central 1 kpc. 
Therefore, it is possible that the AGN outflow influence the gas in the SB ring.
However, investigating this possibility is not straightforward because complex structures and phenomena, such as bar ends and active star formation are intricately involved in the regions where the AGN outflow may interact with the gas in the SB ring. Additionally, the production of each molecule is affected by various phenomena, such as the AGN and star formation. Unfortunately, there are limited established production mechanisms to determine the cause behind the detection of a particular line, but such mechanisms are studied for some molecules. For example, it is known that HCN emission is increased by IR pumping and mechanical heating (e.g., \cite{Imanishi et al. 2009}; \cite{Harada et al. 2010}; \cite{Sakamoto et al. 2010}), while CN enhance due to photodissociation region (PDR) or UV by the AGN outflow (e.g., \cite{Ginard et al. 2015}; \cite{Saito et al. 2022}).
Consequently, the intensity of molecular lines can increase or decrease due to various phenomena. 
Therefore, in analyses such as multi-line modeling that estimate physical quantities from observational data, it is difficult for us to interpret what phenomena are causing these physical quantities to increase or decrease.
For these reasons, it is challenging to determine whether the AGN outflow are indeed affecting the gas in the SB ring.
To address this, we also apply PCA to the region within a diameter of approximately 750 pc -- 2 kpc (10" -- 27".5) (hereafter SB ring region) because PCA has the potential to reconstruct the maps of different phenomena based on the variance in the data space (see section \ref{PCA_outline}). Therefore, it has the potential to visualize solely the effects of the AGN outflow on the SB ring.

In this paper, observations and data processing are described in Section 2. Results of PCA to the overall region (hereafter PCA$_{\rm OA}$) and the SB ring region (hereafter PCA$_{\rm SB}$) are described in Section 3. Discussion of the PCA$_{\rm OA}$ and the PCA$_{\rm SB}$ results is described in Section 4. Summary is in Section 5.

\section{Observations and Processing}
\subsection{ALMA Observations, Ancillary Data, and Data Processing} \label{data processing}

We carried out an imaging spectral scan at 3 mm toward NGC 1068 using Band 3 of ALMA with its 12 m array (Project Code: 2013.1.00279.S, PI: T. Nakajima). Detailed information about these observations is available in \citet{Nakajima et al. 2023}. Additionally, we retrieved and processed all publicly available archival Band 3 data for NGC 1068 as of summer 2021. From this dataset, we selected observations with a spatial resolution of 150 pc or better (angular resolution < 2.08"), resulting in a final sample of eight projects: 2011.0.00061.S, 2012.1.00657.S, 2013.1.00060.S, 2015.1.00960.S, 2017.1.00586.S, 2018.1.01506.S, 2018.1.01684.S, and 2019.1.00130.S, in addition to 2013.1.00279.S. While some of these projects include observations using 7 m array and total power array, we restricted our analysis to data from the 12 m array to ensure consistent maximum recoverable scales across all detected lines. Then, we reconstructed the integrated intensity maps using the PHANGS-ALMA imaging pipeline (\cite{Leroy et al. 2021}). 
We set the grid size to 150 pc because this resolution corresponds to the "cloud-scale" (e.g., \cite{Leroy et al. 2021b}). Additionally, we choose the hexagonal grid to minimize the inter-pixel correlation.

Here is a brief explanation of the above data processing.
We applied observatory-delivered calibration with minor manual data flagging using the appropriate CASA version. 
We subtracted the continuum in the $uv$-plane by fitting a first-degree polynomial function using the CASA uvcontsub task. 
We performed imaging and deconvolution using CLEAN. While the standard PHANGS-ALMA pipeline applies both single-scale CLEAN and multiscale CLEAN, we applied only single-scale CLEAN in this study. 
Next, we smoothed the cube data to a spatial resolution of 150 pc. 
Afterward, we created the integrated intensity map by integrating over the range of $\pm$300 km/s from 1116 km/s, which is the systemic velocity of NGC 1068 (\cite{GarcíaBurillo et al. 2014}). Finally, we regridded the data into a hexagonal grid to minimize correlations between grids. These operations were applied to all molecular lines using the same procedure. For more details, refer to \citet{Leroy et al. 2021}, \citet{Saito et al. 2022a} and \citet{Saito et al. 2022}.

\begin{figure*}[htbp]
\includegraphics[width=17cm]{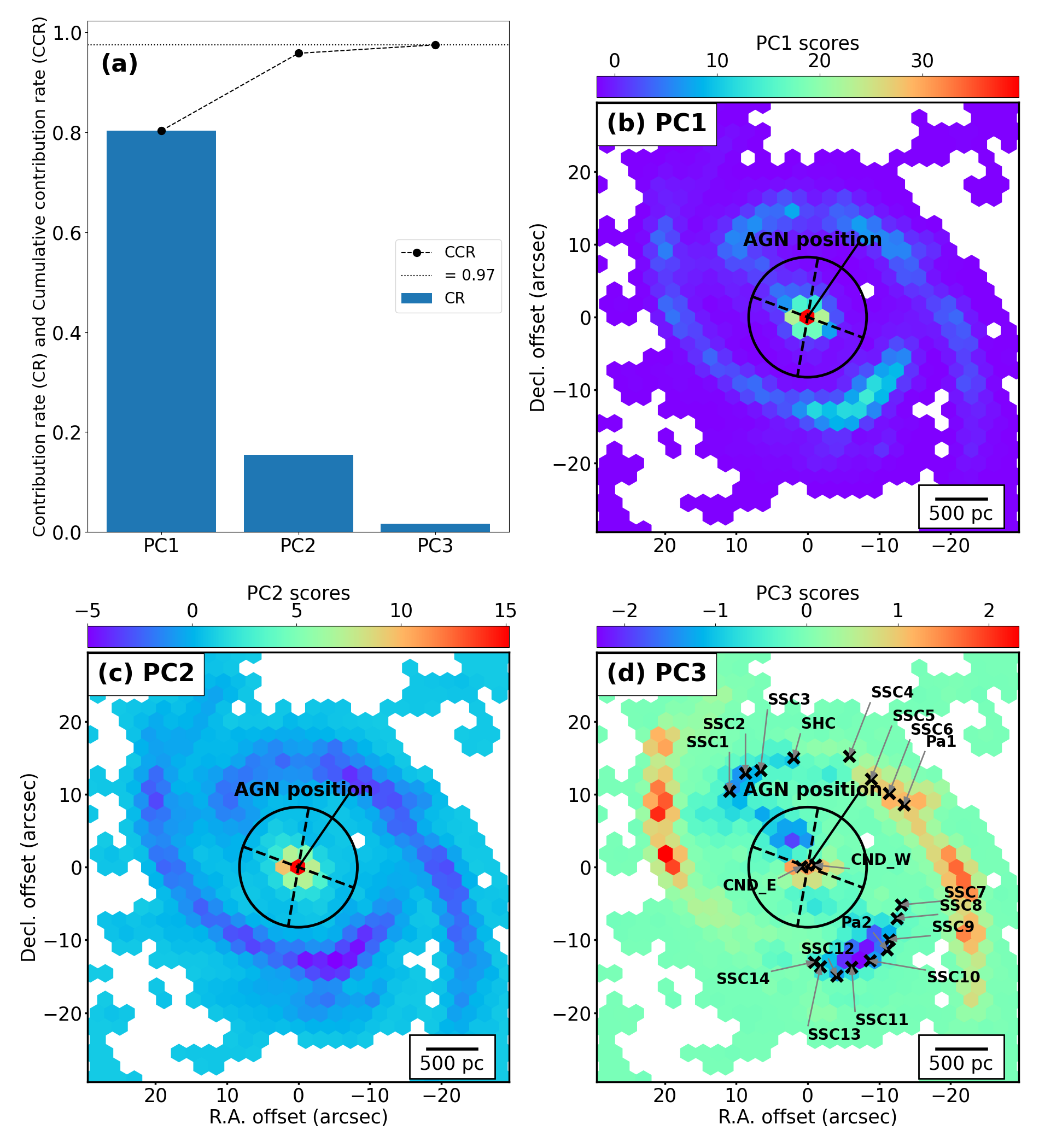}
\caption{
PCA results in overall region: (a) contribution rate (CR) of each PC and cumulative contribution rate (CCR), (b) PC1 score map, (c) PC2 score map, (d) PC3 score map. The AGN position, the two dashed lines and the black circle are same as Figure \ref{13co_maps}. The crosses points in PC3 score map indicate super star clusters (SSCs) (\cite{Rico-Villas et al. 2021}).\\
{Alt text: The results applied PCA for overall region labeled from (a) to (d). (a) shows contribution rate and cumulative contribution rate. From (b) to (d) show PC1 score map, PC2 score map and PC3 score map for overall region.}
}
\label{pca_whole}
\end{figure*}

\begin{figure*}[htbp]
\centering
\includegraphics[width=18cm]{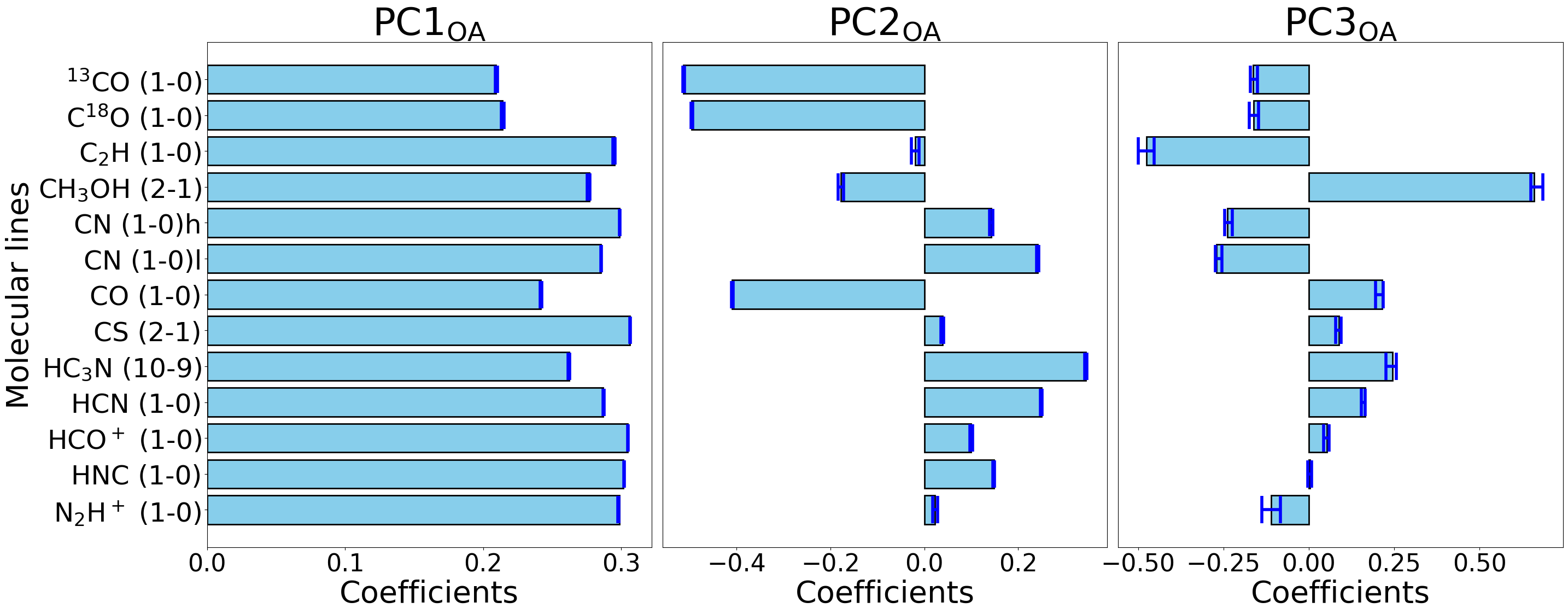}
\caption{PCA$_{\rm OA}$ coefficients.\\
{Alt text: The three subfigures by PCA. From left to right, they represent PC1 coefficients, PC2 coefficients, and PC3 coefficients for overall region.}
}
\label{coefficient_whole}
\end{figure*}

After we remove the grids from all molecular integrated intensity maps where the S/N of the CO ($J$=1-0) integrated intensity map is $\leq$ 4. This step helps minimize the adverse effects of extremely diffuse gas regions on the PCA.
Then, we select the molecular lines to be used in the PCA. If all molecular lines were used, the results might be overly influenced by a specific subset of low quality lines, potentially preventing proper feature extraction. Therefore, it is necessary to use molecular lines that are detected across a sufficiently large portion of the map. 
In this study, we set a threshold of 60 grid points where each molecular line has S/N > 4, within the region where CO ($J$=1-0) also shows S/N > 4. This threshold corresponds to approximately one-tenth of all grid points in the overall region.
As a result, 13 molecular lines remain: C$_2$H ($N$=1-0), HCN ($J$=1-0), HCO$^+$ ($J$=1-0), HNC ($J$=1-0), HC$_3$N ($J$=10-9), N$_2$H$^+$ ($J$=1-0), CH$_3$OH ($J$=2$_{\it K}$-1$_{\it K}$), CS ($J$=2-1), C$^{18}$O ($J$=1-0), $^{13}$CO ($J$=1-0), CN ($N$=1$_{3/2}$-0$_{1/2}$), CN ($N$=1$_{1/2}$-0$_{1/2}$) and CO ($J$=1-0). 
We also apply these integrated intensity maps to the PCA$_{\rm SB}$.
To enable a fair comparison across different molecular lines, we apply a process called standardization to the integrated intensity maps of molecular lines in both the overall region and the SB ring region. For each molecular line, we subtract the mean and divide by the standard deviation of its integrated intensity. This results in all molecular lines having a mean of 0 and a standard deviation of 1.


The integrated intensity map of $^{13}$CO ($J$=1-0) created through the above process is shown in Figure \ref{13co_maps}. Figure \ref{13co_maps}(a) represents the integrated intensity map used for PCA$_{\rm OA}$, while Figure \ref{13co_maps}(b) corresponds to the one used for PCA$_{\rm SB}$. 
Note that the values of grids in Figure \ref{13co_maps}(a) and Figure \ref{13co_maps}(b) are different because we scale using standardization for integrated intensity maps of overall region and SB ring region, respectively.

The integrated intensity maps of other molecular lines are also shown in Figure \ref{mom0_whole_1}, \ref{mom0_whole_2}, \ref{mom0_sbr_1} and \ref{mom0_sbr_2}. The detailed in formation of these line maps are summarized in Table \ref{line table}.
The final data set is (13, 683) for the overall region, where the first number is the number of the available maps and the second is the number of the hexagons in each map. The data set of the SB ring region is (13, 482).

\subsection{Principal Component Analysis} \label{PCA_outline}
PCA is an unsupervised machine learning method that extracts important information by creating new axes in the directions of the greatest variance (\cite{Pearson 1901}). 
In this study, integrated intensity maps of two-dimensional array data are transformed into one-dimensional arrays and subjected to PCA under the assumption that each grid represents a single molecular cloud, with no consideration of inter-grid correlations.

Figure \ref{PCA_outline map} illustrates an example image of PCA in three dimensions. Each dimension is the standardized intensities of HCN ($J$=1-0), CH$_3$OH ($J$=2$_{\it K}$-1$_{\it K}$), and CN ($N$=1$_{3/2}$-0$_{1/2}$) shown in Figures \ref{mom0_whole_1} and \ref{mom0_whole_2}, with each sample represented by grids. PC1 axis represents the direction of the greatest variance in the samples, while PC2 axis is orthogonal to PC1 axis and represents the next greatest variance. 
The value of each sample evaluated along the PC$i$ axis is referred to as the PC$i$ score. In this study, we refer to the map reconstructed by the PC$i$ score as the PC$i$ score map.
The relationship between a PC$i$ axis and each integrated intensity is described by the PC$i$ coefficients, which are the weights used to form PC$i$ axis. 
For example, a high PC$i$ coefficient for a particular molecular line means that the integrated intensity map of that molecular line exhibits a correlation similar to the PC$i$ score map. While it is also possible that the molecular line simply has a high integrated intensity, this effect is minimized in this study due to the application of scaling by standardization.

We used the PCA implementation from the scikit-learn project (\cite{Pedregosa et al. 2011}).

\subsection{Evaluation of error in PCA analysis}
We evaluate the error in our PCA analysis by Monte Carlo method.
First, random values are drawn from a normal distribution with the error moment 0 map as the standard deviation ($\sigma$), and these values are added to the original integrated intensity map. This process is repeated for every grid and for all intensity maps. Then, PCA is performed using the modified data. This procedure is repeated 100,000 times. Finally, for all the PCA coefficients, the values corresponding to ±1$\sigma$ are taken as the error bars.

\begin{figure*}[htbp]
\centering
\includegraphics[width=17cm]{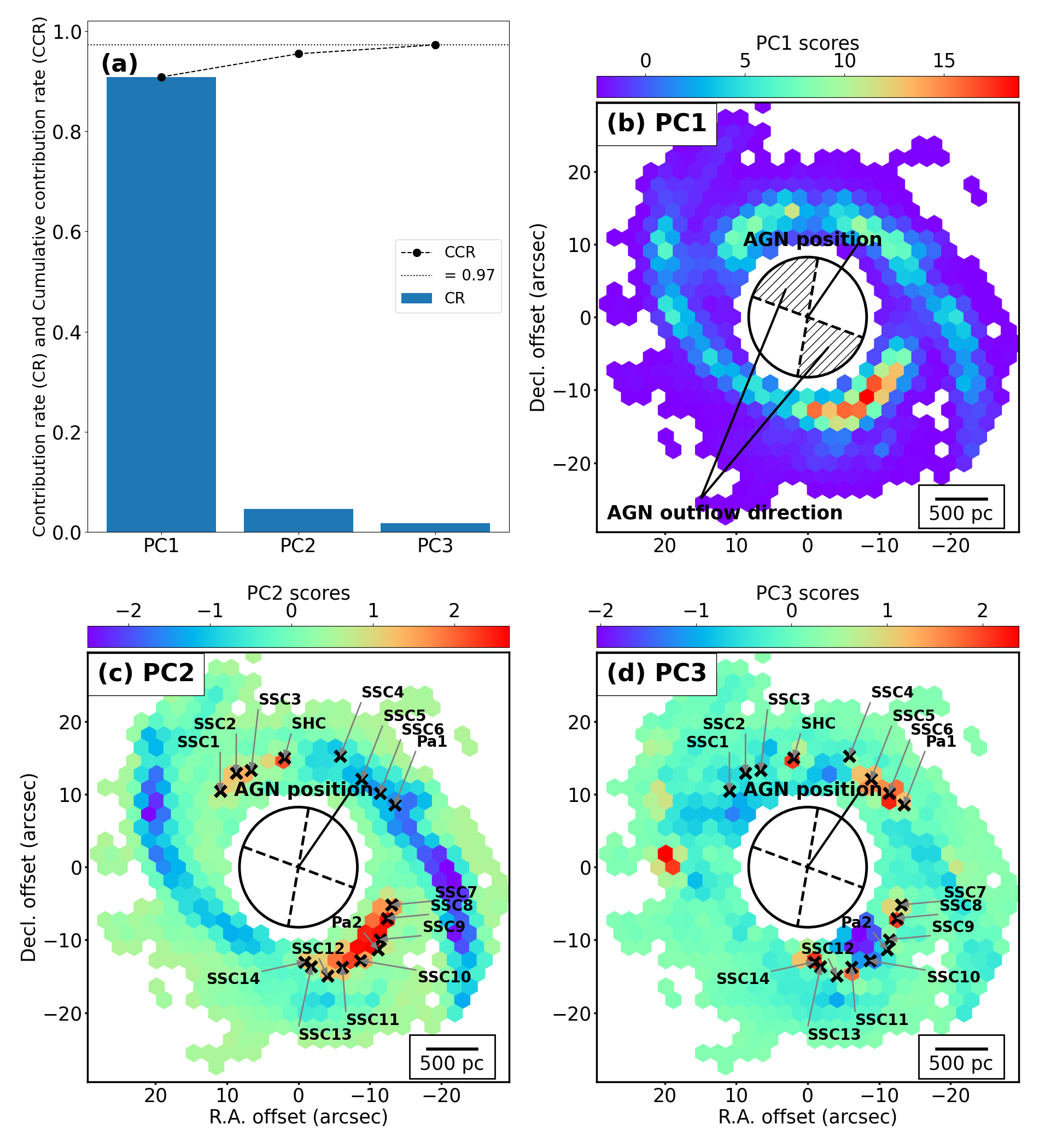}
\caption{
These show PCA results in SB ring region: (a) contribution rate (CR) of each PC and cumulative contribution rate (CCR), (b) PC1 score map, (c) PC2 score map, (d) PC3 score map. The AGN position, the AGN outflow direction, the two dashed lines, the black circle and the crosses points are same as Figure \ref{pca_whole}.\\
{Alt text: The results applied PCA for starburst ring region labeled from (a) to (d). (a) shows contribution rate and cumulative contribution rate. From (b) to (d) show PC1 score map, PC2 score map and PC3 score map for starburst ring region.}
}
\label{pca_sbr}
\end{figure*}

\begin{figure*}[htbp]
\centering
\includegraphics[width=18cm]{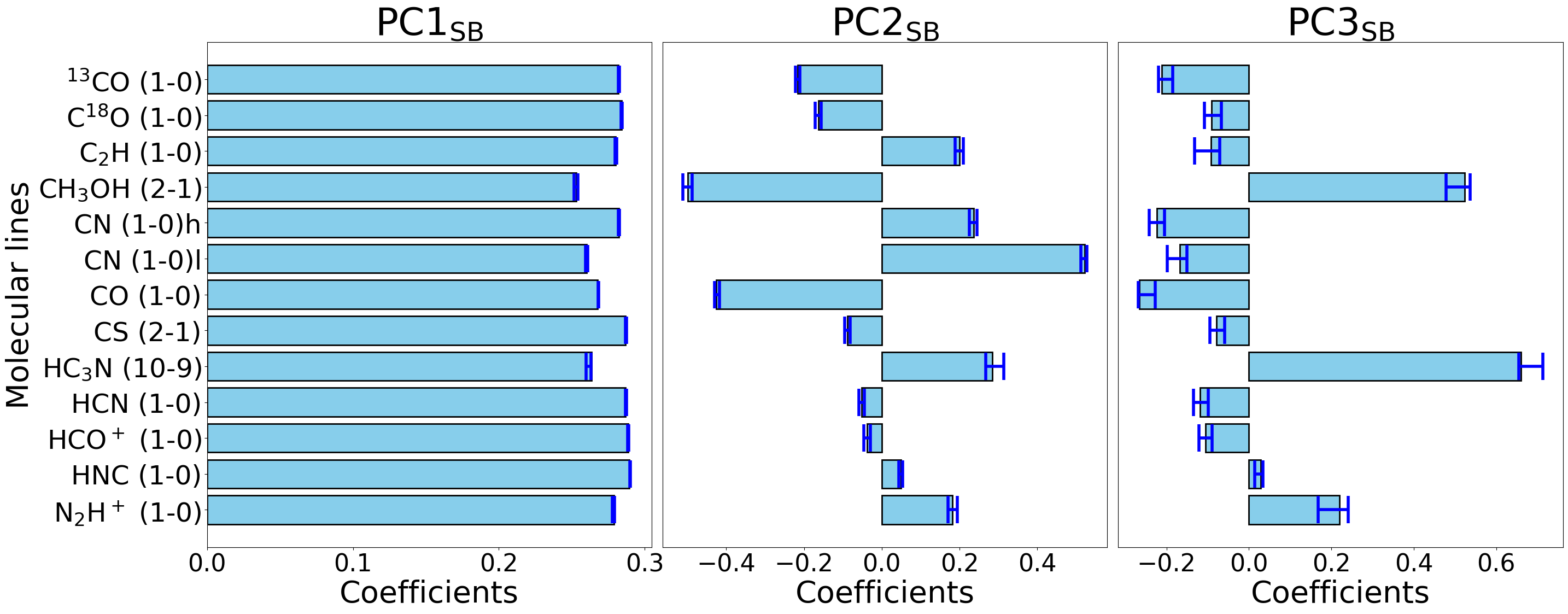}
\caption{PCA$_{\rm SB}$ coefficients.\\
{Alt text: The three subfigures by PCA. From left to right, they represent PC1 coefficients, PC2 coefficients, and PC3 coefficients for starburst ring region.}
}
\label{coefficient_sbr}
\end{figure*}

\section{Results}
Figure \ref{pca_whole} and Figure \ref{coefficient_whole} show the PCA$_{\rm OA}$ results and Figure \ref{pca_sbr} and Figure \ref{coefficient_sbr} show the PCA$_{\rm SB}$ results. Originally, the orientation of PCA axes is arbitrary. In this study, we determined the axes such that the maximum absolute value of the PCA scores is positive.

In this study, we apply PCA to standardized integrated intensity maps of molecular lines to account for variations. For reference, the results using normalization, which is another common scaling method and sets the minimum and maximum values of each molecular line to 0 and 1, are provided in the Appendix. Although the PCA score maps show some visual differences, they are generally minor and do not significantly alter the interpretation.

\renewcommand{\arraystretch}{1.4} 
\begin{table*}[h]
\centering
\caption{Contribution ratio (CR), cumulative contribution ratio (CCR), and coefficients}
\begin{tabular}{l|rrr|rrr}
   \hline
       & \multicolumn{3}{c|}{Overall Region} & \multicolumn{3}{c}{SB Ring Region} \\ \hline
       & PC1$_{\rm OA}$ & PC2$_{\rm OA}$ & PC3$_{\rm OA}$ & PC1$_{\rm SB}$ & PC2$_{\rm SB}$ & PC3$_{\rm SB}$ \\ \hline
   CR  & 0.80 & 0.16 & 0.02 & 0.90 & 0.05 & 0.02 \\
   CCR & 0.80 & 0.96 & 0.97 & 0.90 & 0.95 & 0.97 \\
   \hline\hline
   $^{13}$CO ($J$=1-0) & $0.21_{-0.00065}^{+0.00093}$ & $-0.51_{-0.0018}^{+0.0018}$ & $-0.16_{-0.0082}^{+0.012}$ 
                        & $0.28_{-0.00010}^{+0.00067}$ & $-0.22_{-0.0052}^{+0.0060}$ & $-0.21_{-0.082}^{+0.027}$ \\
                        
   C$^{18}$O ($J$=1-0) & $0.21_{-0.0010}^{+0.00091}$ & $-0.49_{-0.0020}^{+0.0021}$ & $-0.16_{-0.012}^{+0.015}$ 
                        & $0.28_{-0.00015}^{+0.00059}$ & $-0.16_{-0.0078}^{+0.0079}$ & $-0.090_{-0.017}^{+0.023}$ \\
                        
   C$_2$H ($N$=1-0)    & $0.30_{-0.0013}^{+0.00033}$ & $-0.020_{-0.0084}^{+0.0082}$ & $-0.48_{-0.022}^{+0.023}$ 
                        & $0.28_{-0.00061}^{+0.00050}$ & $0.20_{-0.012}^{+0.010}$ & $-0.092_{-0.040}^{+0.022}$ \\
                        
   CH$_3$OH ($J$=2-1)  & $0.28_{-0.0016}^{+0.00034}$ & $-0.18_{-0.0060}^{+0.0061}$ & $0.66_{-0.0089}^{+0.025}$ 
                        & $0.25_{-0.0013}^{+0.0011}$ & $-0.50_{-0.010}^{+0.012}$ & $0.52_{-0.044}^{+0.012}$ \\
                        
   CN ($N$=1$_{3/2}$-0$_{1/2}$)  & $0.30_{-0.00021}^{+0.00041}$ & $0.14_{-0.0035}^{+0.0033}$ & $-0.24_{-0.0079}^{+0.016}$ 
                                & $0.28_{-0.00038}^{+0.00027}$ & $0.24_{-0.010}^{+0.0085}$ & $-0.22_{-0.020}^{+0.019}$ \\
                                
   CN ($N$=1$_{1/2}$-0$_{1/2}$)  & $0.29_{-0.00012}^{+0.00059}$ & $0.24_{-0.0027}^{+0.0026}$ & $-0.27_{-0.0040}^{+0.016}$ 
                                & $0.26_{-0.00052}^{+0.00061}$ & $0.52_{-0.010}^{+0.0061}$ & $-0.17_{-0.031}^{+0.017}$ \\
                                
   CO ($J$=1-0)        & $0.24_{-0.00021}^{+0.00083}$ & $-0.41_{-0.0018}^{+0.0018}$ & $0.21_{-0.020}^{+0.0029}$ 
                        & $0.27_{-0.00015}^{+0.00076}$ & $-0.43_{-0.0042}^{+0.0074}$ & $-0.26_{-0.038}^{+0.0020}$ \\
                        
   CS ($J$=2-1)        & $0.31_{-0.00012}^{+0.00036}$ & $0.038_{-0.0028}^{+0.0025}$ & $0.089_{-0.010}^{+0.0060}$ 
                        & $0.29_{-0.00012}^{+0.00050}$ & $-0.088_{-0.0074}^{+0.0072}$ & $-0.078_{-0.016}^{+0.020}$ \\
                        
   HC$_3$N ($J$=10-9)  & $0.26_{-0.00079}^{+0.00065}$ & $0.34_{-0.0026}^{+0.0018}$ & $0.24_{-0.0019}^{+0.012}$ 
                        & $0.26_{-0.0037}^{+0.000026}$ & $0.28_{-0.0018}^{+0.030}$ & $0.66_{-0.0050}^{+0.052}$ \\
                        
   HCN ($J$=1-0)       & $0.29_{-0.000060}^{+0.00068}$ & $0.25_{-0.0019}^{+0.0032}$ & $0.16_{-0.011}^{+0.00091}$ 
                        & $0.29_{-0.000020}^{+0.00057}$ & $-0.051_{-0.0075}^{+0.0069}$ & $-0.11_{-0.016}^{+0.020}$ \\
                        
   HCO$^+$ ($J$=1-0)   & $0.30_{-0.000090}^{+0.00042}$ & $0.10_{-0.0032}^{+0.0030}$ & $0.053_{-0.0098}^{+0.0059}$ 
                        & $0.29_{-0.00015}^{+0.00043}$ & $-0.037_{-0.0086}^{+0.0077}$ & $-0.11_{-0.015}^{+0.017}$ \\
                        
   HNC ($J$=1-0)       & $0.30_{-0.00018}^{+0.00060}$ & $0.15_{-0.0019}^{+0.0019}$ & $0.0030_{-0.0052}^{+0.0046}$ 
                        & $0.29_{-0.00013}^{+0.00060}$ & $0.049_{-0.0049}^{+0.0039}$ & $0.029_{-0.015}^{+0.0058}$ \\
                        
   N$_2$H$^+$ ($J$=1-0) & $0.30_{-0.0010}^{+0.00014}$ & $0.023_{-0.0053}^{+0.0052}$ & $-0.11_{-0.027}^{+0.027}$ 
                        & $0.28_{-0.00098}^{+0.00039}$ & $0.18_{-0.012}^{+0.013}$ & $0.22_{-0.052}^{+0.020}$ \\
   \hline
\end{tabular}
\label{coefficient_table}
\end{table*}

\subsection{Overall region} \label{PCA_OA}
Figure \ref{pca_whole}(a) shows the contribution rate (CR) and cumulative contribution rate (CCR) based on the integrated intensities of 13 molecular lines. CR indicates how much information each PCA axis retains, while CCR shows the total retained up to that PCA axis. 
The PC1$_{\rm OA}$ score map retains about 80\% of the information, PC2$_{\rm OA}$ score map retains approximately 15\%, and PC3$_{\rm OA}$ score map retains around 2\%. We will focus on these first three principal components since they explain more than 97\% of the total information.
Figure \ref{pca_whole}(b), (c), and (d) show the PCA$_{\rm OA}$ score maps. The PCA$_{\rm OA}$ coefficients are provided in Table \ref{coefficient_table} and Figure \ref{coefficient_whole}.

Figure \ref{pca_whole}(b) shows the PC1$_{\rm OA}$ score map, which looks like features that closely resemble the appearance of most integrated intensity maps. Additionally, the PC1$_{\rm OA}$ coefficients in Figure \ref{coefficient_whole} and Table \ref{coefficient_table} show that all coefficients are positive, even though they differ slightly.

Figure \ref{pca_whole}(c) shows the PC2$_{\rm OA}$ score map, with positive scores near the AGN and the CND, and negative scores along the SB ring. The PC2$_{\rm OA}$ coefficients in Figure \ref{coefficient_whole} and Table \ref{coefficient_table} reveal that HC$_3$N ($J$=10-9), HCN ($J$=1-0), HCO$^+$ ($J$=1-0), HNC($J$=1-0), CN ($N$=1$_{1/2}$-0$_{1/2}$) and CN ($N$=1$_{3/2}$-0$_{1/2}$) contribute most to the positive scores. 
The negative scores are mainly contributed by CO isotopologues. These molecules have weak integrated intensity in the center of NGC 1068.

Figure \ref{pca_whole}(d) shows the PC3$_{\rm OA}$ score map, with positive scores in the CND and the spiral arm, and negative scores in the AGN outflow and the bar-end regions. 
In this paper, we do not identify the spiral arm and the bar-end in detail. We define the spiral arm PC2$_{\rm SB}$ positive scores and the bar-end PC2$_{\rm SB}$ negative scores later in Figure \ref{pca_sbr}(c).
When looking at the PC3$_{\rm OA}$ coefficients in Figure \ref{coefficient_whole} and Table \ref{coefficient_table}, the contribution of CH$_3$OH ($J$=2$_K$-1$_K$) is the largest in the positive scores, followed by HC$_3$N ($J$=10-9), HCN ($J$=1-0), and CO ($J$=1-0). In the negative scores, the contribution of C$_2$H ($N$=1-0) is the largest, followed by CN ($N$=1-0), $^{13}$CO ($J$=1-0), C$^{18}$O ($J$=1-0), and N$_2$H$^+$ ($J$=1-0).

\subsection{SB ring region} \label{SB_ring region}
Figure \ref{pca_sbr}(a) shows the CR and CCR for the SB ring region. PC1$_{\rm SB}$, PC2$_{\rm SB}$, and PC3$_{\rm SB}$ retain nearly 90\%, 5\%, and 2\% of the information, respectively. We focus on them since these three principal components explain about 97\% of the information. The values of the coefficients for each principal component are provided in Table \ref{coefficient_table} and Figure \ref{coefficient_sbr}.

Figure \ref{pca_sbr}(b) represents the PC1$_{\rm SB}$ score map, that resembles the SB ring of the input molecular line maps, with high scores near the southwestern part of the SB ring. The PC1$_{\rm SB}$ coefficients in Figure \ref{coefficient_sbr} and Table \ref{coefficient_table} show that all coefficient values for PC1$_{\rm SB}$ are positive.

Figure \ref{pca_sbr}(c) shows the PC2$_{\rm SB}$ score map, with positive scores located near the bar-end and negative scores near the spiral arms. The PC2$_{\rm SB}$ coefficients in Figure \ref{coefficient_sbr} and Table \ref{coefficient_table} indicate that HC$_3$N ($J$=10-9), N$_2$H$^+$ ($J$=1-0), C$_2$H ($N$=1-0), CN ($N$=1$_{1/2}$-0$_{1/2}$) and CN ($N$=1$_{3/2}$-0$_{1/2}$) contribute most of the positive scores. 
For negative scores, CH$_3$OH ($J$=2$_K$-1$_K$)  and CO ($J$=1-0) are the major contributors.

Figure \ref{pca_sbr}(d) shows the PC3$_{\rm SB}$ score map. Positive scores are located in the eastern spiral arms and certain SSCs. Negative scores are located in the AGN outflow direction. 
The PC3$_{\rm SB}$ coefficients in Figure \ref{coefficient_sbr} and Table \ref{coefficient_table} show that HC$_3$N ($J$=10-9), N$_2$H$^+$ ($J$=1-0) and CH$_3$OH ($J$=2$_K$-1$_K$) contributes most to positive scores. For negative scores, CN ($N$=1$_{1/2}$-0$_{1/2}$), CN ($N$=1$_{3/2}$-0$_{1/2}$), C$_2$H ($N$=1-0), HCN ($J$=1-0), HCO$^+$ ($J$=1-0), CS ($J$=2-1) and CO isotopologues are the major contributors.


\begin{table}[h]
\caption{Tracer Types and Corresponding Molecules}
\begin{tabular}{l|l}
   \hline
   \textbf{Tracer Type} & \textbf{Molecules} \\ \hline
   Molecular gas tracer   & CO ($J$=1-0), $^{13}$CO ($J$=1-0), \\
                        & C$^{18}$O ($J$=1-0) \\ 
   Dense gas tracer     & C$_2$H ($N$=1-0), HCN ($J$=1-0), \\ 
                         & HNC ($J$=1-0), HC$_3$N ($J$=10-9), \\
                         & N$_2$H$^+$ ($J$=1-0), HCO$^+$ ($J$=1-0), \\
                         & CS ($J$=2-1), CN ($N$=1$_{3/2}$-0$_{1/2}$), \\
                         & CN ($N$=1$_{1/2}$-0$_{1/2}$) \\
   Enhancement by shock         & CH$_3$OH \\
   Enhancement by UV            & CN, C$_2$H \\
   Enhancement \\
    by high-temperature  & HCN, HC$_3$N \\
   \hline
\end{tabular}
\label{tab:tracers}
\end{table}

\section{Discussion}
In this section, we discuss based on the PCA coefficients.
For the purpose of the following discussion, Table \ref{tab:tracers} lists the representative roles of the molecular lines used in the PCA as tracers.
CO isotopologues trace molecular gas. The effects of radiative trapping need to be considered because CO ($J$=1-0) is generally optically thick. As a result, CO ($J$=1-0) has a slightly lower critical density compared to $^{13}$CO ($J$=1-0) and C$^{18}$O ($J$=1-0).
C$_2$H ($N$=1-0), HCN ($J$=1-0), HCO$^+$ ($J$=1-0), HNC ($J$=1-0), HC$_3$N ($J$=10-9), N$_2$H$^+$ ($J$=1-0), CS ($J$=2-1), CN ($N$=1$_{3/2}$-0$_{1/2}$) and CN ($N$=1$_{1/2}$-0$_{1/2}$) are tracers of dense gas.
CH$_3$OH is generally considered a shock tracer in external galaxies. (e.g., \cite{Watanabe et al. 2003}; \cite{Flower & Pineau des Forets 2012}; \cite{Saito et al. 2017}).
CN and C$_2$H are thought to be enhanced by UV radiation such as PDR and the AGN outflow (e.g., \cite{GarcíaBurillo et al. 2017}; \cite{Saito et al. 2022}).
HCN and HC$_3$N are thought to be enhanced by high-temperature in the CND (e.g., \cite{Harada et al. 2013}).
A more detailed description of the molecular lines is provided in the following sections.

\subsection{Overall region}
\subsubsection{PC1$_{\rm OA}$} \label{PC1_OA}


According to Section \ref{PCA_OA}, the PC1$_{\rm OA}$ score map in Figure \ref{pca_whole}(b) contains approximately 80\% of the information from all integrated intensity maps and extracts features that closely resemble the appearance of most integrated intensity maps. Additionally, the PC1$_{\rm OA}$ coefficients reveals that all molecular lines contribute to the PC1$_{\rm OA}$ scores. 

Based on these results, the PC1$_{\rm OA}$ scores can be interpreted as extracting regions where gas is present to the extent that molecular lines are excited, suggesting that it reflects the approximate H$_2$ column density.
In fact, \citet{Harada et al. 2024} revealed a strong correlation between PC1 and the H$_2$ column density. Thus, our result do not contradict these previous studies.

However, it should be noted that PCA is inherently a linear combination of the input data, meaning that PC1$_{\rm OA}$ does not precisely extract the H$_2$ column density in detail. This applies not only to the PC1$_{\rm OA}$ but to all PCA$_{\rm OA}$ and PCA$_{\rm SB}$ results.

\subsubsection{PC2$_{\rm OA}$}

\begin{table*}[h]
\centering
\caption{The correlation coefficient between PC2$_{\rm OA}$ scores and the integrated intensity maps of molecules detected ALMA Band3 not used in the PCA$_{\rm OA}$.}
\begin{tabular}{l|l|l|l|l|l|l|l|l|l}
   \hline
   & C$^{17}$O & CH$_3$CN & H$^{13}$CN & H$^{13}$CO$^+$ & HC$_3$N & HC$_3$N & HNCO & SiO & SO \\ 
  & ($J$=1-0) & ($J$=5$_K$-4$_K$) & ($J$=1-0) & ($J$=1-0) & ($J$=11-10) & ($J$=12-11) & ($J$=4$_{0,4}$-3$_{0,3}$) & ($J$=2-1) & ($J$=3$_2$-2$_1$)
        \\ \hline
   Positive scores  & 0.00 &0.75 & 0.96 & 0.93 & 0.96 & 0.96 & 0.39 & 0.95 & 0.95 \\ 
   \hline
   Negative scores  & -0.66 & -0.27 & -0.38 & 0.00 & -0.48 & -0.41 & -0.49 & 0.00 & -0.48 \\
   \hline
\end{tabular}
\label{The correlation coefficient with PC2_OA}
\end{table*}

According to Section \ref{PCA_OA}, the PC2$_{\rm OA}$ score map in Figure \ref{pca_whole}(c) contains approximately 15\% of the information from all integrated intensity maps. The positive scores extract features corresponding to the CND and AGN regions, while the negative scores capture features similar to those of the SB ring.

Examining the PC2$_{\rm OA}$ coefficients in Figure \ref{coefficient_whole} and Table \ref{coefficient_table} reveals that the positive scores are dominated by contributions from highly dense gas tracers such as HC$_3$N ($J$=10-9) and HCN ($J$=1-0), whereas the negative scores are mainly influenced by CO isotopologues, which trace molecular gas (Table \ref{tab:tracers}). Based on these results, PC2$_{\rm OA}$ scores can be interpreted as capturing differences in gas density. \citet{Scourfield et al. 2020} reported that the CND region has a higher density than the SB ring. Therefore, the PC2$_{\rm OA}$ results do not contradict this study of \citet{Scourfield et al. 2020}.

\begin{table*}[h]
\centering
\caption{The correlation coefficient between the PC3$_{\rm SB}$ negative scores and the velocity dispersion with negative PC3$_{\rm SB}$ coefficients}
\begin{tabular}{l|l|l|l|l|l|l|l|l|l}
   \hline
              & $^{13}$CO & C$^{18}$O & C$_2$H & CN & CN & HCN & HCO$^+$ & CS & CO \\ 
              & ($J$=1-0) & ($J$=1-0) & ($N$=1$_{1/2}$-0$_{1/2}$) & ($N$=1$_{3/2}$-0$_{1/2}$) & ($N$=1$_{1/2}$-0$_{1/2}$) & ($J$=1-0) & ($J$=1-0) & ($J$=2-1) & ($J$=1-0)\\ \hline

    Negative & -0.51 & -0.26 & -0.50 & -0.65 & -0.49 & -0.56 & -0.57 & -0.58 & -0.49 \\
    scores  &  &  &  &  &  &  &  &  &  \\
   \hline
\end{tabular}
\label{The correlation coefficient with mom2}
\end{table*}

Additionally, differences in chemical composition may also play a role in the interpretation of the PC2$_{\rm OA}$. 
When examining the positive coefficients in Figure \ref{coefficient_whole}, we understand that coefficients of HCN ($J$=1-0) and HC$_3$N ($J$=10-9) are high. 
These molecules are enhanced at high-temperatures in the CND (Table \ref{tab:tracers}). Coefficients of CN ($N$=1$_{1/2}$-0$_{3/2}$), and HNC ($J$=1-0) are also high. These molecules are considered to trace an X-ray Dominated Region (XDR) (\cite{Lepp and Dalgarno 1996}; \cite{Meijerink and Spaans 2005}).
In fact, The CND, represented by the positive scores, is thought to be XDR and high-temperatures (\cite{GarcíaBurillo et al. 2010}; \cite{Harada et al. 2013}). 
However, HC$_3$N is also expected to be destroyed by X-rays and UV radiation (e.g., \cite{Harada et al. 2013}). \citet{Harada et al. 2013} suggest the presence of a hot midplane in the CND of NGC 1068, which could shield against X-rays and UV. Thus, the result of our study does not contradict the existence of a hot midplane. 

In CND region, the contribution of CO isotopologues is relatively low. So it might seem contradictory to the shielded midplane because CO isotopologues are typically destroyed by UV radiation. However, observations of CO ($J$=3-2) and $^{13}$CO ($J$=3-2) have high intensity near the CND (\cite{Tsai et al. 2012}; \cite{Nakajima et al. 2015}). This suggests that the intensities of CO ($J$=1-0) and $^{13}$CO ($J$=1-0) are lower near the CND due to the high excitation temperature.
Hence, positive scores may represent changes in the chemical composition associated with the complex phenomena in the CND.
We also examine the correlation coefficient between PC2$_{\rm OA}$ scores and the integrated intensity maps of molecules detected ALMA Band3 not used in the PCA$_{\rm OA}$. These integrated intensity maps have undergone the same data processing as described in section \ref{data processing}. Additionally, we use only integrated intensity maps with more than 5 grids satisfying S/N > 4 because calculating the correlation coefficient requires a certain number of samples. The results are shown in Talbe \ref{The correlation coefficient with PC2_OA}. Standardized integrated intensities of CH$_3$CN ($J$=5$_K$-4$_K$), H$^{13}$CN ($J$=1-0), H$^{13}$CO$^+$ ($J$=1-0), HC$_3$N ($J$=12-11), HC$_3$N ($J$=11-10), SiO ($J$=2-1) and SO ($J$=3$_2$-2$_1$) showed a particularly strong correlation with PC2$_{\rm OA}$ positive scores. CH$_3$CN and SO are thought to be destroyed by UV radiation, while SO is enhanced by X-rays (\cite{Aladro et al. 2013}). 
Therefore, we consider that CH$_3$CN enhance in the shielded midplane, while SO enhance in the XDR. 
Moreover, H$^{13}$CN is considered to be enhanced by mechanical heating, such as turbulence and shocks, while SiO is thought to be enhanced by strong shock such as AGN feedback (e.g., \cite{Nakajima et al. 2018}). Such complex phenomena imply that the CND undergoes significant chemical enrichment, as suggested by \citet{Saito et al. 2022}.

For the negative scores, in addition to the strong contributions from CO isotopologues, CH$_3$OH also has a contribution. Furthermore, Table \ref{The correlation coefficient with PC2_OA} shows that, in addition to C$^{17}$O ($J$=1–0), HNCO ($J$=4–3) also has a negative correlation with negative scores (r $\sim$ –0.5). CH$_3$OH and HNCO are considered shock tracers (e.g., \cite{Garay and Lizano. 1999}; \cite{Watanabe et al. 2003}; \cite{Rodr\'{\i}guez-Fern\'{a}ndez et al. 2010}; \cite{Xie et al. 2023}). Furthermore, \citet{Tosaki et al. 2017} reported that shocks caused by interactions between molecular clouds occur in the spiral arms of NGC 1068. Therefore, the PC2$_{\rm OA}$ results do not contradict the study of \citet{Tosaki et al. 2017}.


\subsubsection{PC3$_{\rm OA}$} \label{PC3_OA}

According to Section \ref{PCA_OA}, the PC3$_{\rm OA}$ score map in Figure \ref{pca_whole}(d) contains approximately 2\% of the total information. The positive scores correspond to features resembling the spiral arms and the CND, while the negative scores resemble the bar-end and AGN outflow. Unlike PC1$_{\rm OA}$ and PC2$_{\rm OA}$, PC3$_{\rm OA}$ appears to extract multiple distinct features.

Examining the PC3$_{\rm OA}$ coefficients in Figure \ref{coefficient_whole} and Table \ref{coefficient_table} reveals that the positive scores are influenced by shock tracers such as CH$_3$OH ($J$=2$_K$-1$_K$), molecular gas tracers like CO ($J$=1-0), and highly dense gas tracers including HC$_3$N ($J$=10-9), HCN ($J$=1-0), and CS ($J$=2-1) (Table \ref{tab:tracers}). HCN and HC$_3$N are considered to be enhanced in the high-temperature regions within the CND (Table \ref{tab:tracers}). On the other hand, the negative scores are dominated by species that are enhanced by UV radiation, such as C$_2$H, CN, as well as highly dense gas tracers like N$_2$H$^+$ ($J$=1-0) and molecular gas tracers such as $^{13}$CO ($J$=1-0) and C$^{18}$O ($J$=1-0) (Table \ref{tab:tracers}). These results support the interpretation that the PC3$_{\rm OA}$ score map extracts multiple pieces of information.

\citet{Tosaki et al. 2017} reported that shocks between molecular clouds occur in the spiral arms of NGC 1068. The contributions of CH$_3$OH ($J$=2$_K$-1$_K$) and CO ($J$=1-0) to the positive scores suggest that they may reflect collisions between molecular gas in the spiral arms, consistent with the interpretation of \citet{Tosaki et al. 2017}. Furthermore, \citet{Scourfield et al. 2020} reported that the CND is in a high-density state, and \citet{Harada et al. 2013} indicated that the CND is also in a high-temperature state. Additionally, \citet{Huang and Viti 2023} suggested that CH$_3$OH may be enhanced in the CND due to its high-temperature. Therefore, the significant contributions of HC$_3$N, HCN, CS, and CH$_3$OH to the positive scores can be interpreted as reflecting the high-temperature and high-density conditions in the CND, which do not contradict those previous studies. Based on these PC3$_{\rm OA}$ results, the positive scores may be extracting information related to molecular cloud collisions in the spiral arms, and the high-temperature and high-density state in the CND.

\citet{Saito et al. 2022} reported that UV radiation from the AGN outflow in NGC 1068 enhances CN and C$_2$H.
The contributions of CN and C$_2$H to the negative scores suggest that they correspond to the negative scores associated with the AGN outflow, consistent with the interpretation of \citet{Saito et al. 2022}. Additionally, the bar-end of this galaxy exhibit higher star formation rate surface densities compared to other SB ring regions (\cite{Tsai et al. 2012}). Consequently, high-density gas, which serves as the material for star formation, is expected to be abundant in the bar-end. Since $^{13}$CO ($J$=1-0) and C$^{18}$O ($J$=1-0) have higher critical densities compared to CO ($J$=1-0), the contributions of $^{13}$CO ($J$=1-0), C$^{18}$O ($J$=1-0), CN ($N$=1$_{3/2}$-0$_{1/2}$), CN ($N$=1$_{1/2}$-0$_{1/2}$), C$_2$H ($N$=1-0), and N$_2$H$^+$ ($J$=1-0) to the negative scores may be interpreted as extracting intermediate to high-density gas in the bar-end. This interpretation do not contradict \citet{Tsai et al. 2012}.

\subsection{SB ring region}

\subsubsection{PC1$_{\rm SB}$}
According to Section \ref{SB_ring region}, the PC1$_{\rm SB}$ score map in Figure \ref{pca_sbr}(b) contains approximately 90\% of the total information and primarily extracts features resembling the integrated intensity maps of molecular lines. Furthermore, examining the PC1$_{\rm SB}$ coefficients in Figure \ref{coefficient_sbr} and Table \ref{coefficient_table} reveals that all molecular lines contribute to the PC1$_{\rm SB}$. This suggests that, consistent with previous studies such as \citet{Meier and Turner 2005} and \citet{Harada et al. 2024}, the PC1 scores broadly reflect the H$_2$ column density. However, as noted in Section \ref{PC1_OA}, it is important to keep in mind that PC1$_{\rm SB}$ does not represent the H$_2$ column density in a detailed manner.

\subsubsection{PC2$_{\rm SB}$}


According to Section \ref{SB_ring region}, the PC2$_{\rm SB}$ score map in Figure \ref{pca_sbr}(c) contains approximately 5\% of the total information, with positive scores highlighting the bar-end region and negative scores extracting features resembling the spiral arms.

Examining the PC2$_{\rm SB}$ coefficients in Figure \ref{coefficient_sbr} and Table \ref{coefficient_table} reveals that the positive scores are primarily influenced by HC$_3$N ($J$=10-9) and N$_2$H$^+$ ($J$=1-0), which trace highly dense gas, as well as CN and C$_2$H, which are enhanced by UV radiation such as PDR and dense gas tracers (Table \ref{tab:tracers}). Furthermore, the positive scores spatially coincide with most SSCs.
These suggest that the PC2$_{\rm SB}$ positive scores extract regions of active star formation.
Indeed, these regions exhibit higher star formation rate (SFR) surface densities compared to other areas (e.g., \cite{Tsai et al. 2012}). Therefore, the positive scores do not contradict these previous studies. However, it should be noted that the positive scores do not perfectly coincide with the locations of SSCs since PCA does not allow for a detailed extraction of regions with active star formation.

Additionally, the PC2$_{\rm SB}$ coefficients in Figure \ref{coefficient_sbr} and Table \ref{coefficient_table} indicate that the negative scores are mainly influenced by CO ($J$=1-0), which traces molecular gas, and CH$_3$OH ($J$=2$_K$-1$_K$), which is associated with shocks. This interpretation aligns with the positive scores in the spiral arm regions of PC3$_{\rm OA}$, suggesting that PC2$_{\rm SB}$ negative scores reflect collisions between molecular molecular clouds. 
This result do not contradict \citet{Tosaki et al. 2017}, which reported molecular cloud shocks in the spiral arms of NGC 1068.


\subsubsection{PC3$_{\rm SB}$}

\begin{figure*}[htbp]
\centering
\includegraphics[width=18cm]{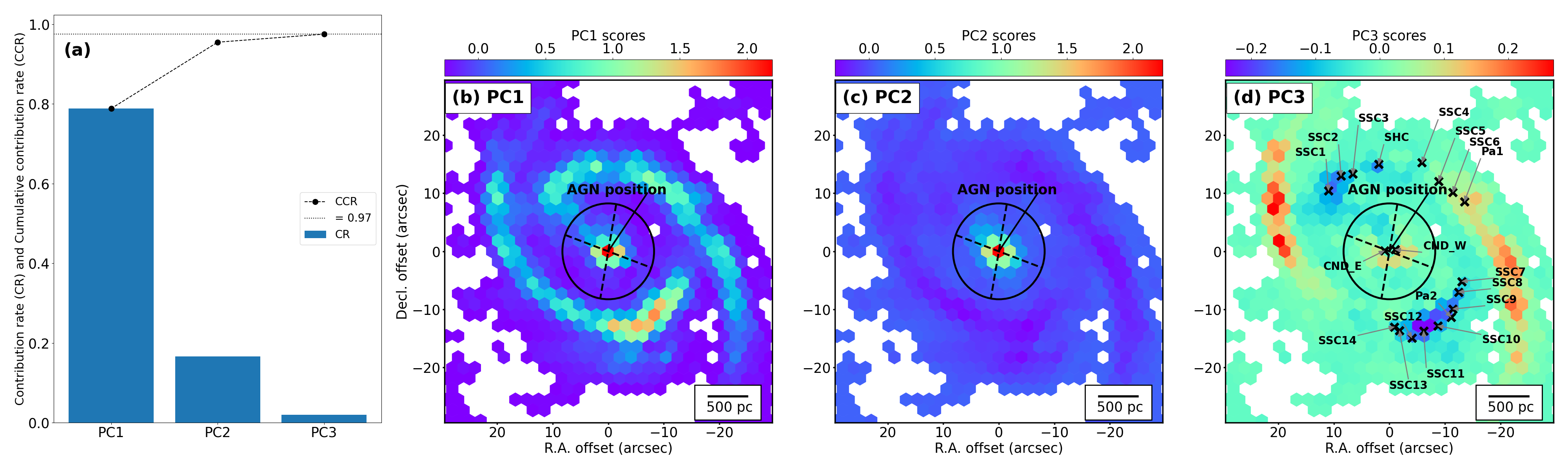}
\caption{The PCA$_{\rm OA}$ results by normalized integrated intensity maps.\\
{Alt text: The results applied PCA for overall region by normalized integrated intensity maps labeled from (a) to (d). (a) shows contribution rate and cumulative contribution rate. From (b) to (d) show PC1 score map, PC2 score map and PC3 score map for overall region.}
}
\label{normalized PCA_OA}
\end{figure*}

\begin{figure*}[htbp]
\centering
\includegraphics[width=16cm]{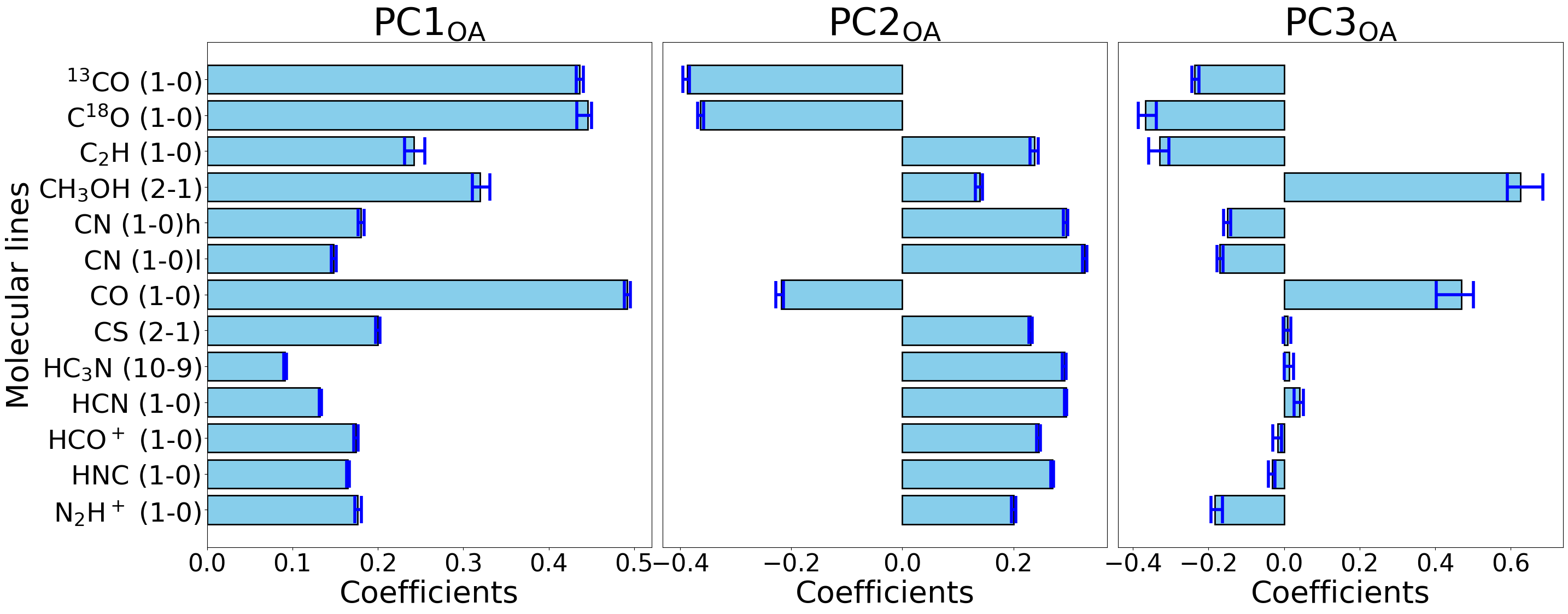}
\caption{The PCA$_{\rm OA}$ coefficients.\\
{Alt text: The three subfigures by PCA by normalized integrated intensity maps. From left to right, they represent PC1 coefficients, PC2 coefficients, and PC3 coefficients for overall region.}
}
\label{normalized PCA_OA coefficients}
\end{figure*}

According to Section \ref{SB_ring region}, the PC3$_{\rm SB}$ score map in Figure \ref{pca_sbr}(d) contains approximately 2\% of the total information, with positive scores located in the eastern spiral arm and certain SSCs, while negative scores are positioned along the AGN outflow direction.

Examining the PC3$_{\rm SB}$ coefficients in Figure \ref{coefficient_sbr} and Table \ref{coefficient_table} reveals that the positive scores are primarily contributed by molecules tracing dense gas, such as HC$_3$N ($J$=10-9) and N$_2$H$^+$ ($J$=1-0), as well as CH$_3$OH ($J$=2$_K$-1$_K$), which is a shock tracer (Table \ref{tab:tracers}). This suggests that the positive scores reflect some information related to star formation. \citet{Rico-Villas et al. 2021} identified the Superhot Core (SHC), SSC5, SSC6, SSC11, and SSC14 as relatively young star clusters ($\sim 10^4 - 10^5$ yr). Additionally, \citet{Nagashima et al. 2024} reported that shock-driven star formation is occurring in the eastern spiral arm. The regions reported in these previous studies roughly coincide with the locations of the positive scores, suggesting that the positive scores capture subtle variations in integrated intensity related to shocks and the relatively young star clusters.

On the other hand, the negative scores are contributed by molecular lines such as CN and C$_2$H, which are enhanced by UV radiation, CO isotopologues tracing molecular gas, and dense gas tracers like HCN ($J$=1-0), HCO$^+$ ($J$=1-0), and CS ($J$=2-1) (Table \ref{tab:tracers}). It is likely that the negative scores are extracting multiple pieces of information because a consistent interpretation of the negative coefficients is difficult. No particular SSCs exhibit distinctly strong negative scores, making it unlikely that the negative scores directly reflect star formation activity. \citet{Nakajima et al. 2023} reported that the AGN outflow may extend beyond a diameter of 1 kpc (indicated by the circle in Figure \ref{pca_sbr}(d)).
Therefore, while the negative scores may contain multiple pieces of information, one possible interpretation of which could be the interaction between the AGN outflow and the gas in the SB ring.
Previous studies have reported that CN and C$_2$H can be enhanced by UV radiation from AGN outflow (\cite{Saito et al. 2022}), and HCN can be enhanced either by shock or high-temperature induced by the AGN outflow (e.g., \cite{Aalto et al. 2012}; \cite{Izumi et al. 2013}). Additionally, if the gas in the SB ring is being compressed by the AGN outflow, it is consistent that dense gas tracers such as HCN ($J$=1-0), HCO$^+$ ($J$=1-0) and CS ($J$=2-1) contribute to the negative scores. Therefore, we cannot rule out the possibility that one of the pieces of information captured by the negative scores represents the interaction between the AGN outflow and the gas in the SB ring. 

If the AGN outflow interacts with gas in the SB ring, the velocity dispersion  of molecular lines can be larger. Thus, the PC3$_{\rm SB}$ negative scores and velocity dispersion should have anti-correlation. We examine the correlation between the velocity dispersion of $^{13}$CO ($J$=1-0), C$^{18}$O ($J$=1-0), C$_2$H ($N$=1$_{1/2}$-0$_{1/2}$), CN ($N$=1$_{1/2}$-0$_{1/2}$), CN ($N$=1$_{3/2}$-0$_{1/2}$), HCN ($J$=1-0), HCO$^+$ ($J$=1-0), and CS ($J$=2-1), which have high contributions to the PC3$_{\rm SB}$ negative scores, and the negative scores. We note that C$_2$H has relatively large fine and hyperfine structures. 
Therefore, the velocity dispersion of C$_2$H includes the effect of such structure.
This study utilizes only C$_2$H ($N$=1$_{1/2}$-0$_{1/2}$) because the hyperfine splitting within each fine structure line is nearly identical (e.g., \cite{GarcíaBurillo et al. 2017}). The results are shown in Table \ref{The correlation coefficient with mom2}. The negative scores and the velocity dispersions of $^{13}$CO ($J$=1-0), CN ($N$=1$_{3/2}$-0$_{1/2}$), C$_2$H ($N$=1$_{1/2}$-0$_{1/2}$), HCN ($J$=1-0), HCO$^+$ ($J$=1-0), and CS ($J$=2-1) are in anti-correlation (r < -0.5). 
When looking at this, we can see the correlation except for C$^{18}$O ($J$=1-0). 
Therefore, this has a possibility that the PC3$_{\rm SB}$ negative scores likely extract information about shock between the AGN outflow and gas in the SB ring. On the other hand, a more detailed analysis is required because these regions also overlap with the bar-end and have possibility that multiple features are being extracted.


Examining results of the PCA$_{\rm OA}$ and the PCA$_{\rm SB}$, it can be said that up to PC2, the extracted features are relatively easy to interpret. However, PC3 appears to capture multiple pieces of information. This is likely due to the fact that the complex phenomena occurring within the NGC 1068 introduce nonlinearity into the data structure, whereas PCA is a linear feature extraction method. Since the PC3$_{\rm SB}$ suggests a possible interaction between the AGN outflow and the gas in the SB ring, it is necessary to extract low-contribution PC score maps in detail. One potential approach is to employ nonlinear feature extraction methods.

\begin{figure*}[htbp]
\centering
\includegraphics[width=18cm]{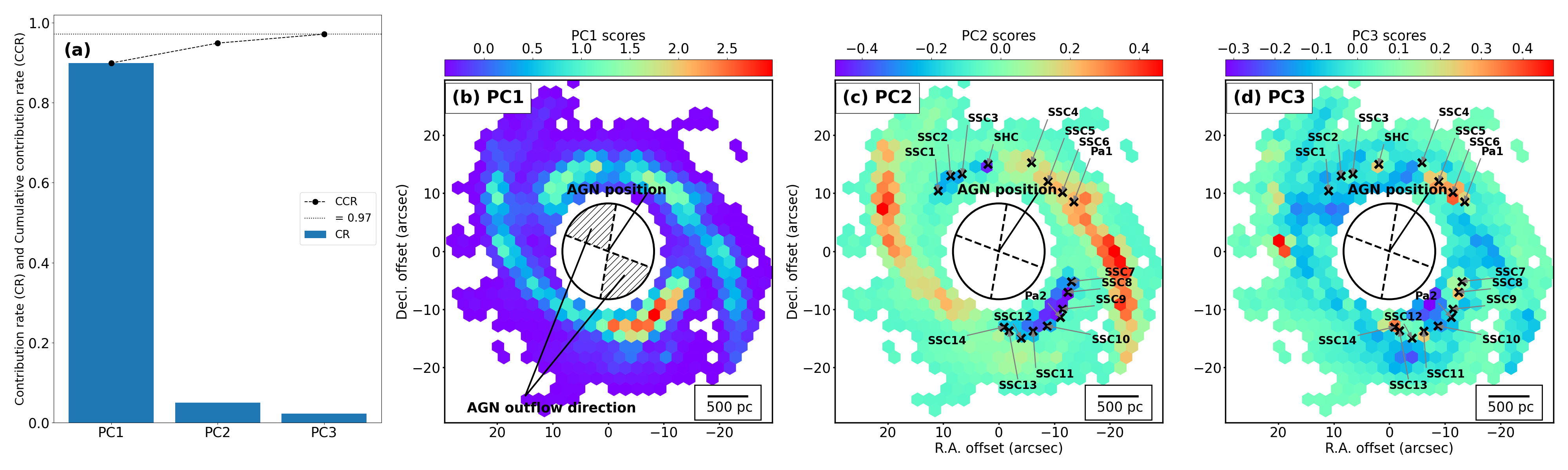}
\caption{The PCA$_{\rm SB}$ results by normalized integrated intensity maps.\\
{Alt text: The results applied PCA for starburst ring region by normalized integrated intensity maps labeled from (a) to (d). (a) shows contribution rate and cumulative contribution rate. From (b) to (d) show PC1 score map, PC2 score map and PC3 score map for starburst ring region.}
}
\label{normalized PCA_SB}
\end{figure*}

\begin{figure*}[htbp]
\centering
\includegraphics[width=16cm]{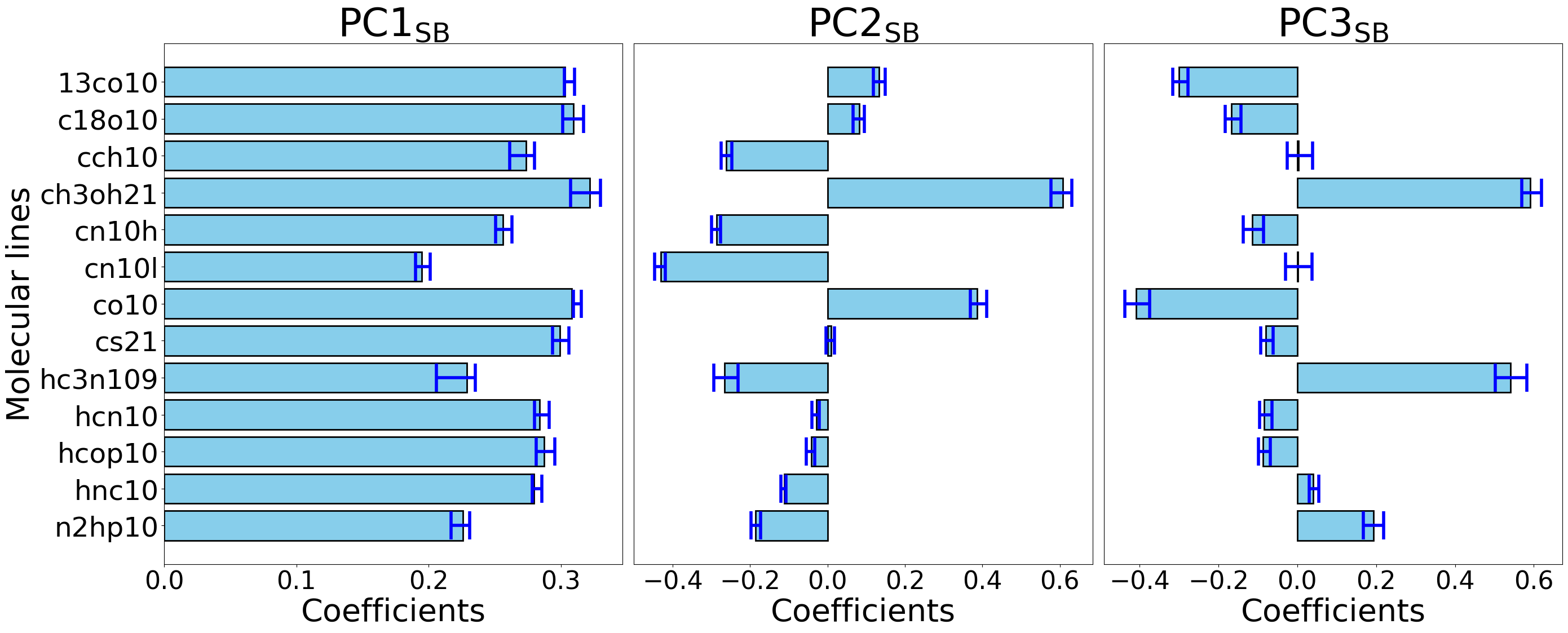}
\caption{The PCA$_{\rm SB}$ coefficients.\\
{Alt text: The three subfigures by PCA by normalized integrated intensity maps. From left to right, they represent PC1 coefficients, PC2 coefficients, and PC3 coefficients for starburst ring region.}
}
\label{normalized PCA_SB coefficients}
\end{figure*}

\section{Summary}
In this paper, we applied PCA to the overall region and the SB ring region in NGC 1068 to extract characteristic features of the molecular lines.

The PCA$_{\rm OA}$ results in (1) extracting the distribution of the entire molecular cloud, (2) extracting the features of the CND and the SB ring and (3) extracting the features of both the CND and the spiral arm and both the AGN outflow and the bar-end.

The PCA$_{\rm SB}$ results in (1) extracting the distribution of the entire molecular cloud, (2) extracting the structures of the bar-end and the spiral arms and (3) extracting features that are thought to be influenced by young star-forming regions and that the AGN outflow is possibly influencing the gas in the SB ring.

This study demonstrated that PCA can effectively extract features even in complex galaxies with the AGN and the SB ring. Additionally, the results of PC3$_{\rm SB}$ imply that examining low-contribution PC maps may help uncover phenomena that are physically challenging to detect. However, further detailed investigation is required because low-contribution PC maps may extract multiple pieces of information simultaneously.

\begin{ack} 

This paper makes use of the following ALMA data: ADS/JAO.ALMA\#2011.0.00061.S, ADS/JAO.ALMA\#2012.1.00657.S, ADS/JAO.ALMA\#2013.1.00060.S, ADS/JAO.ALMA\#2015.1.00960.S, ADS/JAO.ALMA\#2017.1.00586.S, ADS/JAO.ALMA\#2018.1.01506.S, ADS/JAO.ALMA\#2018.1.01684.S, and ADS/JAO.ALMA\#2019.1.00130.S. 

This work was supported by JST SPRING, Grant Number JPMJSP2124. T.S. was supported by the Daiichi-Sankyo “Habataku” Support Program for the Next Generation of Researchers, and by the Japan Foundation for Promotion of Astronomy. This work was supported by National Astronomical Observatory of Japan ALMA Scientific Research
Grant No. 2021-18A.
S.T. acknowledges Nihon University research grant for 2025 Overseas Researchers.
N.H. acknowledges support from JSPS KAKENHI grant No. JP25K07375.
K.M. acknowledges financial support from the Japan Society for the Promotion of Science (JSPS) through KAKENHI grants No. 20K14516 and, 23H00131.

ALMA is a partnership of ESO (representing its member states), NSF (USA) and NINS (Japan), together with NRC (Canada), MOST and ASIAA (Taiwan), and KASI (Republic of Korea), in cooperation with the Republic of Chile. The Joint ALMA Observatory is operated by ESO, AUI/NRAO and NAOJ. This research has made use of the NASA/IPAC Extragalactic Database, which is funded by the National Aeronautics and Space Administration and operated by the California Institute of Technology. The National Radio Astronomy Observatory is a facility of the National Science Foundation operated under a cooperative agreement by Associated Universities, Inc. This research has made use of the SIMBAD database, operated at CDS, Strasbourg, France. Data analysis was in part carried out on the Multi-wavelength Data Analysis System operated by the Astronomy Data Center, National Astronomical Observatory of Japan.

{\it Software}: ALMA Calibration Pipeline, Astropy (\cite{Astropy Collaboration et al. 2013}; \cite{Astropy Collaboration et al. 2018}), CASA (\cite{McMullin et al. 2007}), NumPy (\cite{Harris et al. 2020}), PHANGS-ALMA Pipeline (\cite{Leroy et al. 2021}), SciPy (\cite{Virtanen et al. 2020}), spectral-cube (\cite{Ginsburg et al. 2019}).

\end{ack}


\appendix
\section*{The PCA results by the normalized integrated intensity maps}
In the main analysis, we apply standardization to scale each molecular line such that their variances are unified. As an alternative approach, this appendix presents the results of PCA performed on the molecular line data scaled by normalization, which transforms all values to a range between 0 and 1.

Figure \ref{normalized PCA_OA} and \ref{normalized PCA_OA coefficients} presents the PCA$_{\rm OA}$ score maps and PCA$_{\rm OA}$ coefficients, respectively. 
While the PC1$_{\rm OA}$ score map and the PC2$_{\rm OA}$ score map in Figure \ref{normalized PCA_OA}(b) and (c) differ from those derived from standardized molecular line data in Figure \ref{pca_whole}(b) and (c), the differences do not significantly affect the core interpretation.
However, the PC3$_{\rm OA}$ negative score map exhibits notable differences. The negative scores by standardized molecular lines in Figure \ref{pca_whole}(d) are interpreted as extracting the AGN outflow and the star formation and/or PDR at the bar-end region (see Section \ref{PC3_OA}). On the other hand, the negative scores by normalized molecular lines in Figure \ref{normalized PCA_OA}(d) primarily extract only the bar-end region. So when looking at the PC3$_{\rm OA}$ coefficients in Figure\ref{normalized PCA_OA coefficients}, The contributions of $^{13}$CO ($J$=1–0) and C$^{18}$O ($J$=1–0) are dominant, as they trace regions denser than those probed by CO ($J$=1–0), followed by the dense gas tracers  N$_2$H$^+$ ($J$=1-0), and CN and C$_2$H, which are PDR tracers and dense gas tracers.

Figure \ref{normalized PCA_SB} and \ref{normalized PCA_SB coefficients} presents the PCA$_{\rm SB}$ score maps and the PCA$_{\rm SB}$ coefficients, respectively. 
Similar to the PCA$_{\rm OA}$ results, while the PC1$_{\rm SB}$ score map and the PC2$_{\rm SB}$ score map in Figure \ref{normalized PCA_SB}(b) and (c) differ from those derived from standardized molecular line data in Figure \ref{pca_sbr}(b) and (c), the differences do not significantly affect the core interpretation.
However, the PC3$_{\rm SB}$ negative score map exhibits notable differences.
The negative scores in Figure \ref{pca_sbr} (d) show the highest negative scores in the southwestern region in the SB ring, while lower negative scores appear in the northeast. On the other hand, the difference in negative scores between the southwestern and northeastern regions of the SB ring becomes less pronounced in Figure \ref{normalized PCA_SB}(d).
Examining the PC3$_{\rm SB}$ coefficients in Figure \ref{normalized PCA_SB coefficients} reveals that CO isotopologues have the largest contribution, followed by CN ($N$=1$_{3/2}$-0$_{1/2}$). This suggests that the negative scores in Figure \ref{normalized PCA_SB}(d) contain more contributions from diffuse gas compared to the negative scores in Figure \ref{pca_sbr}(d).


\end{document}